\newif{\ifchangetext}
\changetextfalse

\documentclass[iop]{emulateapj}
\newcommand{\cmt}[1]{}
\ifchangetext
  \newcommand{\change}[1]{{\bf \textcolor{blue}{#1}}}
  \newcommand{\changeG}[1]{#1}
\else
  \newcommand{\change}[1]{#1}
  \newcommand{\changeG}[1]{#1}
\fi

\usepackage{natbib}   
\usepackage{amsmath}  
\usepackage{verbatim} 
\usepackage{enumerate}
\usepackage{amssymb}  
\usepackage{hyperref}      
\usepackage[usenames,dvipsnames]{xcolor}  
\usepackage{appendix}

\DeclareGraphicsExtensions{.pdf,.png,.jpg}

\defcitealias{Scolnic:2009}{S09}
\def\S09{\citetalias{Scolnic:2009}}

\defcitealias{Rodney:2014}{R14}
\def\R14{\citetalias{Rodney:2014}}

\def\ie{{i.e.}}

\def\Om{\ensuremath{\Omega_{\rm m}}}

\def\LCDM{$\Lambda$CDM}

\def\Ho{\ensuremath{H_0}}

\def\arcsec{\ensuremath{^{\prime\prime}}}

\def\Msun{\mbox{M$_{\odot}$}}
\def\Av{\mbox{$A_V$}}
\def\Rv{\mbox{$R_V$}}

\newcommand{\bush}{GSD11Bus}
\newcommand{\colfax}{GND12Col}
\newcommand{\stone}{GND13Sto}

\newcommand{\CCSN}{CC\,SN}
\newcommand{\CCSNe}{CC\,SNe}

\newcommand{\SNIa}{Type Ia SN}
\newcommand{\SNeIa}{Type Ia SNe}

\def\Spitzer{{\it Spitzer}}

\shorttitle{SN Ia at $z\sim2$ in HST Medium Bands}
\shortauthors{Rodney et al.}

\begin{document}

\title{Two Type Ia Supernovae at Redshift $\sim$2 : 
Improved Classification and Redshift Determination 
with Medium-band Infrared Imaging }

\newcommand{\HubbleFellow}{Hubble Fellow}
\newcommand{\JHU}{Department of Physics and Astronomy, The Johns Hopkins University, 3400 N. Charles St., Baltimore, MD 21218, USA}
\newcommand{\USC}{Department of Physics and Astronomy, University of South Carolina, 712 Main St., Columbia, SC 29208, USA}
\newcommand{\elsewhere}{St. Elsewhere University.}
\newcommand{\STScI}{Space Telescope Science Institute, 3700 San Martin Dr., Baltimore, MD 21218, USA}
\newcommand{\Chicago}{Department of Physics, The University of Chicago, Chicago, IL 60637, USA}
\newcommand{\Berkeley}{Department of Astronomy, University of California, Berkeley, CA 94720-3411, USA}
\newcommand{\LBNL}{Lawrence Berkeley National Laboratory, Berkeley, CA 94720, USA}
\newcommand{\Riverside}{Department of Physics and Astronomy, University of California, Riverside, CA 92521, USA}
\newcommand{\SaoPaulo}{Instituto de Astronomia, Geof\'isica e Ci\^encias
Atmosf\'ericas, Universidade de S\~ao Paulo, Cidade Universit\'aria,
05508-090, S\~ao Paulo, Brazil}
\newcommand{\Andalucia}{Instituto de Astrof\'isica de Andaluc\'ia (CSIC), E-18080 Granada, Spain}
\newcommand{\Cantabria}{IFCA, Instituto de F\'isica de Cantabria (UC-CSIC), Av. de Los Castros s/n, 39005 Santander, Spain}
\newcommand{\WKU}{Department of Physics, Western Kentucky University, Bowling Green, KY 42101, USA}
\newcommand{\AMNH}{Department of Astrophysics, American Museum of Natural History, Central Park West and 79th Street, New York, NY 10024, USA}
\newcommand{\NYU}{Center for Cosmology and Particle Physics, New York University, New York, NY 10003, USA}
\newcommand{\Copenhagen}{Dark Cosmology Centre, Niels Bohr Institute, University of Copenhagen, Juliane Maries Vej 30, DK-2100 Copenhagen, Denmark}
\newcommand{\Arizona}{Department of Astronomy, University of Arizona, Tucson, AZ 85721, USA}
\newcommand{\SantaCruz}{Department of Astronomy and Astrophysics, University of California, Santa Cruz, CA 92064, USA}
\newcommand{\NotreDame}{Department of Physics, University of Notre Dame, Notre Dame, IN 46556, USA}
\newcommand{\TelAviv}{Department of Astrophysics, Tel Aviv University, 69978 Tel Aviv, Israel}
\newcommand{\Rutgers}{Department of Physics and Astronomy, Rutgers, The State University of New Jersey, Piscataway, NJ 08854, USA}
\newcommand{\CfA}{Harvard/Smithsonian Center for Astrophysics, Cambridge, MA 02138, USA}
\newcommand{\Minnesota}{Department of Astronomy, University of Minnesota, 116 Church Street SE, Minneapolis, MN 55455, USA}
\newcommand{\NOAO}{National Optical Astronomical Observatory, Tucson, AZ 85719, USA}
\newcommand{\UCSB}{Department of Physics, University of California, Santa Barbara, CA 93106-9530, USA}
\newcommand{\SantaBarbara}{\UCSB}
\newcommand{\LCOGT}{Las Cumbres Observatory Global Telescope Network, 6740 Cortona Dr., Suite 102, Goleta, California 93117, USA}
\newcommand{\Colby}{Colby College, 4000 Mayflower Hill Dr, Waterville, ME 04901, USA}
\newcommand{\Kentucky}{University of Kentucky, Lexington, KY 40506}
\newcommand{\UCDavis}{University of California Davis, 1 Shields Avenue, Davis, CA 95616}

\newcounter{affilct}
\setcounter{affilct}{0}

\makeatletter
\newcommand{\affilref}[1]{%
  \@ifundefined{c@#1}%
    {\newcounter{#1}%
     \setcounter{#1}{\theaffilct}%
     \refstepcounter{affilct}%
     \label{#1}%
     }{}%
  \ref{#1}%
 }
\makeatother

\makeatletter
\newcommand*\affilreftxt[2]{%
  \@ifundefined{c@#1txt}
    {\newcounter{#1txt}%
     \setcounter{#1txt}{1}
     \altaffiltext{\ref{#1}}{#2}
     }{
     }
  }
\makeatother

\author{Steven~A.~Rodney\altaffilmark{\affilref{JHU},\affilref{USC},\affilref{HubbleFellow}}}
\affilreftxt{JHU}{\JHU}
\affilreftxt{USC}{\USC}
\affilreftxt{HubbleFellow}{\HubbleFellow}
\email{srodney@sc.edu}

\author{Adam G.~Riess\altaffilmark{\affilref{JHU},\affilref{STScI}}}
\affilreftxt{JHU}{\JHU}
\affilreftxt{STScI}{\STScI}

\author{Daniel~M.~Scolnic\altaffilmark{\affilref{Chicago}}}
\affilreftxt{Chicago}{\Chicago}

\author{David~O.~Jones\altaffilmark{\affilref{JHU}}}
\affilreftxt{JHU}{\JHU}

\author{Shoubaneh~Hemmati\altaffilmark{\affilref{Riverside}}}
\affilreftxt{Riverside}{\Riverside}

\author{Alberto Molino\altaffilmark{\affilref{Andalucia},\affilref{SaoPaulo}}}
\affilreftxt{Andalucia}{\Andalucia}
\affilreftxt{SaoPaulo}{\SaoPaulo}

\author{Curtis McCully\altaffilmark{\affilref{LCOGT},\affilref{UCSB}}}
\affilreftxt{LCOGT}{\LCOGT}
\affilreftxt{UCSB}{\UCSB}

\author{Bahram~Mobasher\altaffilmark{\affilref{Riverside}}}
\affilreftxt{Riverside}{\Riverside}

\author{Louis-Gregory Strolger\altaffilmark{\affilref{STScI},\affilref{WKU}}}
\affilreftxt{STScI}{\STScI}
\affilreftxt{WKU}{\WKU}

\author{Or Graur\altaffilmark{\affilref{NYU},\affilref{AMNH}}}
\affilreftxt{NYU}{\NYU}
\affilreftxt{AMNH}{\AMNH}

\author{Brian~Hayden\altaffilmark{\affilref{NotreDame},\affilref{LBNL}}}
\affilreftxt{NotreDame}{\NotreDame}
\affilreftxt{LBNL}{\LBNL}

\author{Stefano~Casertano\altaffilmark{\affilref{STScI}}}
\affilreftxt{STScI}{\STScI}

\begin{abstract}
We present two supernovae (SNe) discovered with the {\em Hubble Space
Telescope} (HST) in the Cosmic Assembly Near-infrared Deep
Extragalactic Legacy Survey, an HST multi-cycle treasury
program.  We classify both objects as Type Ia SNe and find
redshifts of $z=1.80\pm0.02$ and $2.26^{+0.02}_{-0.10}$, the latter of
which is the highest redshift Type Ia SN yet seen.  Using light curve
fitting we determine luminosity distances and find that both objects
are consistent with a standard $\Lambda$CDM cosmological model.  These
SNe were observed using the HST Wide Field Camera 3 infrared detector, 
with imaging in both wide- and medium-band filters.  We
demonstrate that the classification and redshift estimates are
significantly improved by the inclusion of single-epoch medium-band
observations.  This medium-band imaging approximates a very low
resolution spectrum ($\lambda/\Delta\lambda\lesssim$100) which can
isolate broad spectral absorption features that differentiate Type Ia
SNe from their most common core collapse cousins.  This medium-band
method is also insensitive to dust extinction and (unlike grism
spectroscopy) it is not affected by contamination from the SN host
galaxy or other nearby sources.  As such, it can provide a more
efficient -- though less precise -- alternative to IR spectroscopy for
high-$z$ SNe.
\end{abstract}


\section{Introduction}\label{sec:Introduction}

Type Ia supernovae (SNe) at redshifts $z>1$ are a valuable tool 
for measuring the expansion history of the universe and for
testing models of \SNIa\ progenitor systems.
Measurements of the \SNIa\ rate at $z>1$ have been used to constrain
\SNIa\ progenitor models through the delay time distribution
test \citep{Poznanski:2007,Graur:2011,Dahlen:2004,Strolger:2006,Dahlen:2008}
and recent extensions of the sample to $z\sim2$ are providing unique
leverage on the fraction of \SNIa\ progenitors that explode promptly
after formation \citep[][hereafter \R14]{Graur:2014,Rodney:2014}.

A subset of these high-$z$ \SNeIa\ with well sampled light curves has
been used to extend luminosity distance measurements into an era when
the universe was decelerating.  \change{A notable early example is SN
1997ff \citep{Gilliland:1999,Riess:2001}, which held the title of the
farthest known \SNIa\ for over a
decade \citep[until][]{Jones:2013}. By reaching into the early
universe, SN 1997ff and other} high-$z$ \SNeIa\ provided a valuable
check against systematic effects that could mimic
dark energy signatures at lower
$z$ \citep{Riess:2000,Riess:2001,Tonry:2003,Riess:2004b}. As the sample of
high-$z$ \SNeIa\ has grown, it has also delivered limits on the
evolution of the equation of state of the universe over cosmic
time \citep{Riess:2007,Suzuki:2012}.

Discovering \SNeIa\ and measuring their rest-frame optical light
curves at $z>1$ requires deep infrared imaging. This imaging data
alone has been widely used for measuring \SNIa\ rates to test SN
progenitor models, aided by an increasingly rigorous suite of
photometric classification
software \citep[e.g.,][]{Poznanski:2007,Rodney:2009,Kessler:2010,Sako:2011}.

However, SN cosmology is still almost exclusively
limited to samples with spectroscopic classifications
(e.g., \citealt{Rest:2014}, \citealt{Betoule:2014}; but
see \citealt{Campbell:2013}).  Spectroscopic observations of SNe in
the range $1<z<1.5$ with ground-based telescopes are feasible, but very
challenging \citep[e.g.,][]{Lidman:2005,Morokuma:2010}.  The task is
more manageable from space, and slitless grism spectroscopy with the
Hubble Space Telescope (HST) Advanced Camera for Surveys (ACS)
has been used to successfully classify 16 \SNeIa\ in this redshift
range (\citealt{Riess:2004a,Riess:2004b,Riess:2007,Graur:2014}, \citetalias{Rodney:2014}).
However, as SNe discoveries are pushed to
redshifts $z>1.5$, such direct
spectroscopic SN classifications become increasingly
costly and less definitive \citep{Rodney:2012,Rubin:2013,Jones:2013}.

In cases where the SN spectrum cannot independently define the SN
class and redshift, we may make use of photometric classifications and
host galaxy redshifts.  However, biases in the derived redshift and
distance are more likely in such cases, due to errors in host galaxy
association or when a spectroscopic redshift for the host galaxy is
unavailable. Host galaxy redshifts are especially problematic for SNe
in the so-called redshift desert ($1.4<z<2.5$), which is precisely the
current frontier for high-$z$ SN searches with HST.  Newer infrared
spectrometers on 8-m class telescopes are able to define precise
spectroscopic redshifts for some SN host galaxies in this redshift
range, but it remains a challenging
endeavor \citep[e.g.,][]{Amanullah:2011,Frederiksen:2012,Frederiksen:2014}.
These biases can also be reduced -- though not wholly eliminated -- by
designing surveys around galaxy clusters, where SNe appearing in
early-type cluster member galaxies have less ambiguity in both class
and redshift \citep{Dawson:2009}.

In Section~\ref{sec:Method} we present an alternative approach for
determining the classification and redshift of SNe at $z>1.4$.  This
method builds on the ``SN Cross Correlation'' imaging technique
of \citet[][hereafter S09]{Scolnic:2009}, extending it to higher
redshifts by using medium-band imaging from the HST Wide Field Camera
3 infrared (WFC3-IR) detector.  Unlike photometric constraints from
pure broad-band imaging, this approach is insensitive to reddening,
and is able to test for specific individual absorption features in the
SN spectral energy distribution (SED).  This medium-band method is
more efficient than slitless spectroscopy with an HST grism, and is
also largely unaffected by contamination from nearby bright sources.

Section~\ref{sec:Observations} describes the discovery and observation
of two Type Ia SNe at $z\sim2$, found in the SN search component of
the Cosmic Assembly Near-infrared Deep Extragalactic Legacy Survey
(CANDELS, \citealt{Koekemoer:2011,Grogin:2011}).  In
Sections~\ref{sec:Stone} and \ref{sec:Colfax} we apply the HST
medium-band imaging technique to help define the classification and
redshift of each SN.  A third high-redshift SN, classified as
a Core Collapse SN (\CCSN), is presented in the Appendix.  
The value of the medium-band imaging method is examined in
Section~\ref{sec:MedbandDiscussion}, and in
Section~\ref{sec:Cosmology} we put these SN discoveries into
the context of current \SNIa\ cosmology.  Finally, our results and
conclusions are briefly summarized in Section~\ref{sec:Summary}.

\section{Pseudo-Spectroscopy with\\ Medium Band Filters}\label{sec:Method}

The SED of a \SNIa\ near maximum light is dominated by a series of
broad absorption features interspersed with pseudo-continuum peaks,
creating a roughly sinusoidal pattern over the rest-frame wavelength
range 3500-6500\,\AA\ \citep[e.g.,][]{Filippenko:1997}.
\S09 introduced the specialized
SuperNovA Cross Correlation (SNACC) filters, a set of custom optical
filters designed to take advantage of this crudely periodic SED shape.
Each filter has 3 pass-bands (``teeth'') that approximately match the
width and spacing of the SNIa spectral features.  \S09 showed that
SNACC filter observations combined with traditional broad band imaging
from the Pan-STARRS1 SN survey could deliver a \SNIa\ sample with 98\%
purity and redshift precision of $\sigma_z=0.01$, all with an
efficiency up to $\sim$30 times better than traditional spectroscopic
follow-up. \change{In addition to the efficiency gains, this comb filter
approach is largely insensitive to dust extinction, and can easily
employ image subtraction techniques to minimize contamination of the
SN signal by the underlying host galaxy light.}

\begin{figure}
\begin{center}
\includegraphics[width=\columnwidth]{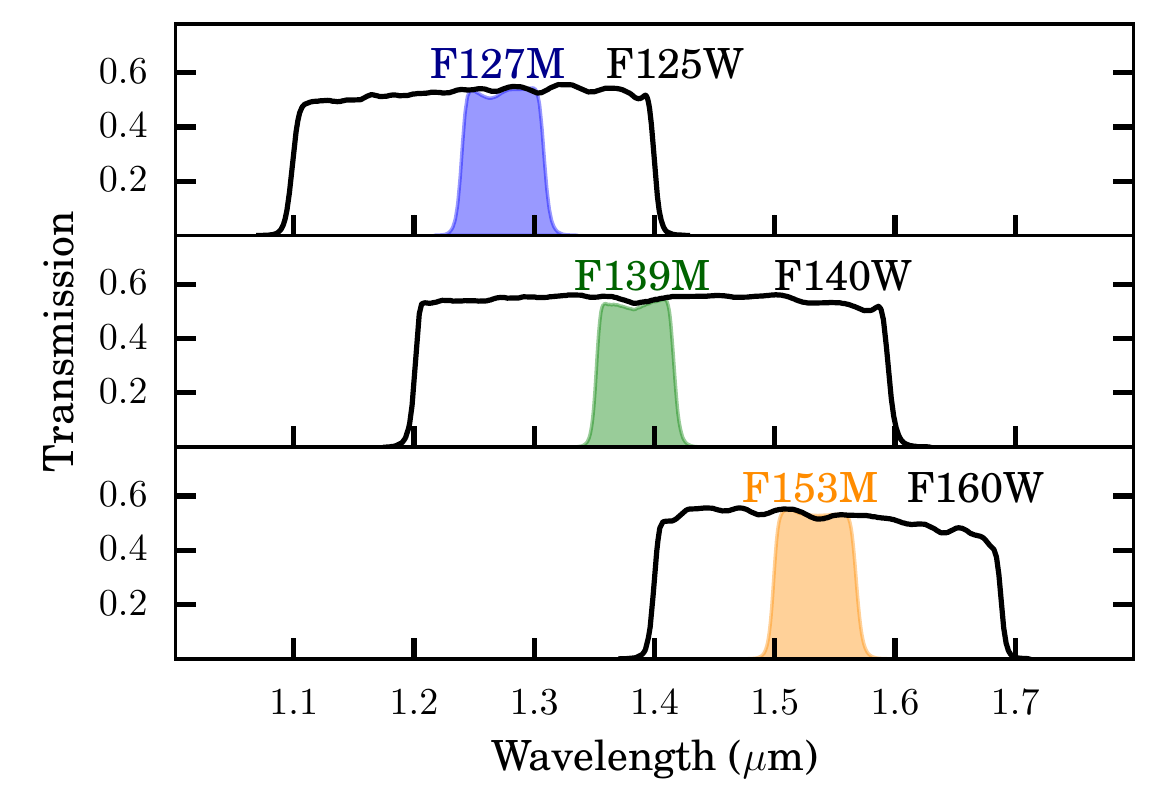}
\caption{
\change{
The three primary medium band filters for the HST WFC3-IR camera, and their complementary wide band filters.  The transmission function of each medium-band filter is shown as a filled curve, and the accompanying wide-band filter as a solid black line.  These three pairs define the medium-minus-wide pseudo-colors used throughout this work. }
\label{fig:Filters} }
\end{center}
\end{figure}

The SNACC filters employed by \S09 were optimized for SNe in the
redshift range $0.2<z<0.8$, in order to target SNe discovered in
ground-based surveys.  To extend this technique to higher redshift
SNe, we must turn to infrared (IR) wavelengths.  The HST WFC3-IR
camera includes a set of medium-band filters (F098M,F127M,F139M,F153M)
which can collectively approximate the teeth of the SNACC comb
filters. These four medium-band filters on the WFC3-IR detector were
designed as grism reference bands or to pick out molecular absorption
features in extrasolar planets (e.g., H$_2$O, CH$_4$, NH$_3$).  The
F098M filter spans 157\,nm, while the three redder bands have a width
of $\sim65$\,nm.  By comparison, the wide-band IR filters such as
F125W ($J$) and F160W ($H$) cover $\sim275$\,nm each.  \change{ As
shown in Figure~\ref{fig:Filters}, each of the medium-band filters has
a corresponding wide-band filter that completely contains the
medium-band filter wavelengths.  Throughout this work we use
{\it medium-minus-wide} pseudo-colors, subtracting the magnitude of a
source in the wide band from the medium-band magnitude (e.g.,
F127M-F125W). This is consistent with the convention set in \S09\ for
the SNACC filters.}

\begin{figure*}
\begin{center}
\includegraphics[width=\textwidth]{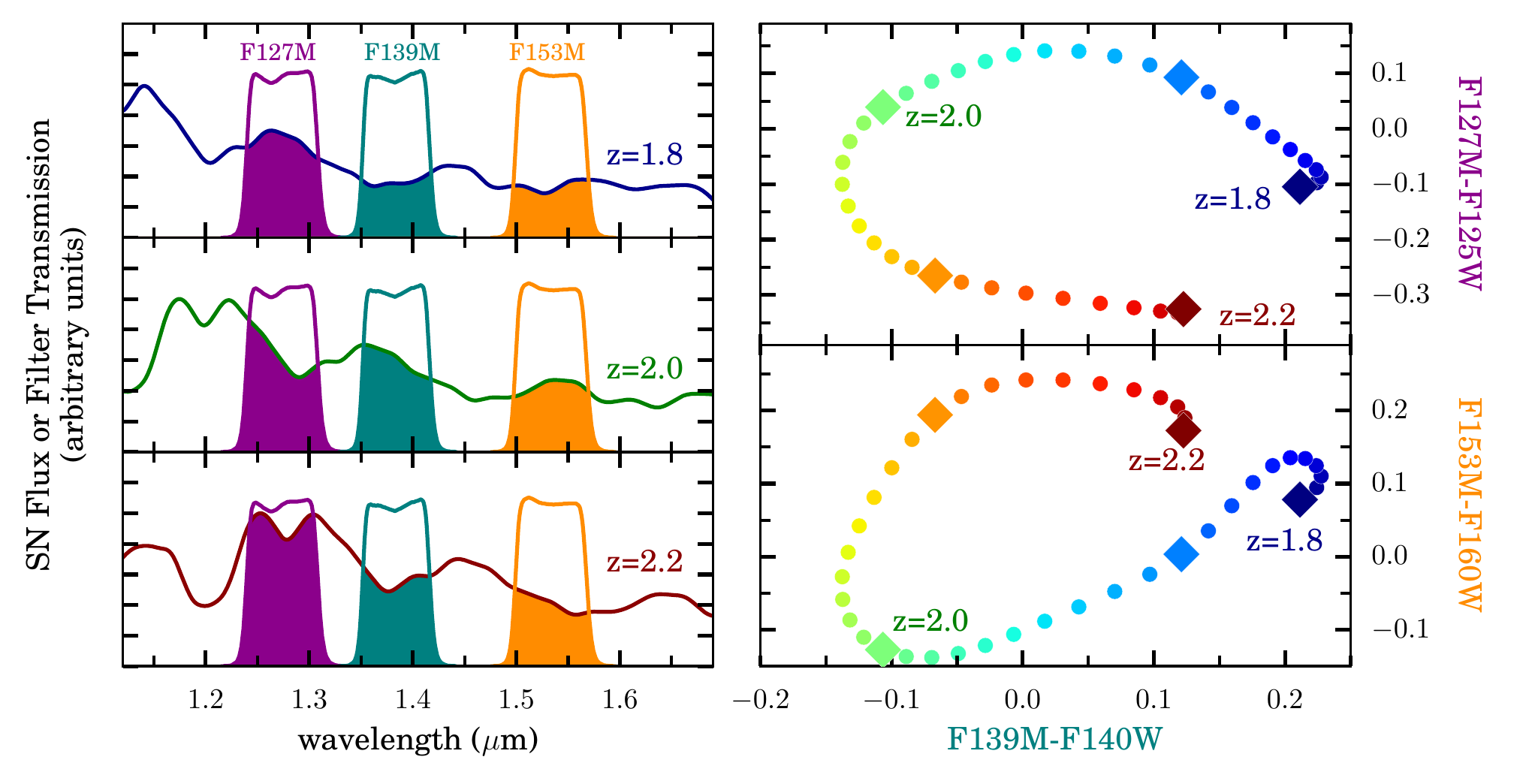}
\caption{  
Demonstration of the method for determining a \SNIa\ redshift from
medium$-$wide band pseudo-colors. Three panels at left show the SED
of a template \SNIa\ \citep{Hsiao:2007} at peak brightness with no extinction.
From top to bottom these panels show the SED at redshifts of 1.8, 2.0,
and 2.2.  Overlaid on each SED are the transmission curves for the
WFC3-IR medium band filters.  As the redshift increases, broad
absorption features are moved in and out of the band passes, causing
cyclic changes in the observed flux through each band, relative to the
encompassing broad bands (not shown).  In two panels at right we plot
the medium$-$wide pseudo-colors, mapping the results of changing the 
redshift from $z=1.8$ to 2.2.  
\label{fig:RedshiftDemo} }
\end{center}
\end{figure*}

The spectral absorption features of \SNeIa\ span roughly 30\,nm in the
rest frame, so for a high-redshift SN at $1.5<z<2.5$ the observer
finds them broadened to widths of $75-105$\,nm.  As shown on the left
side of Figure~\ref{fig:RedshiftDemo}, this means that the medium-band
filter widths are well suited for isolating the spectral features
characteristic of \SNeIa\ in this redshift range. Observations with
traditional wide-band filters must accompany each medium-band
observation in order to provide a continuum reference point.  The
medium-band pseudo-colors (e.g. F139M-F140W) indicate whether the
medium band filter intersects an absorption feature (resulting in a
more positive pseudo-color) or a brighter continuum region (more
negative pseudo-color). As redshift increases, the absorption features
migrate in and out of the medium band regions, so pseudo-colors move
from positive to negative and back again.  In a plot comparing two
pseudo-colors, as shown in Figure~\ref{fig:RedshiftDemo}, a \SNIa\
will trace out a circle as it is moved through redshift space.  For
redshifts $1.5<z<2.5$, measuring three of these pseudo-colors near
peak brightness to better than 0.1 mag can yield a redshift estimate
with precision approaching 2\%\ in $\Delta z/(1+z)$, comparable to the
best available photometric redshifts from galaxy SED
fitting \citep{Hildebrandt:2010,Dahlen:2013}.

\begin{figure*}
\begin{center}
\includegraphics[width=\textwidth]{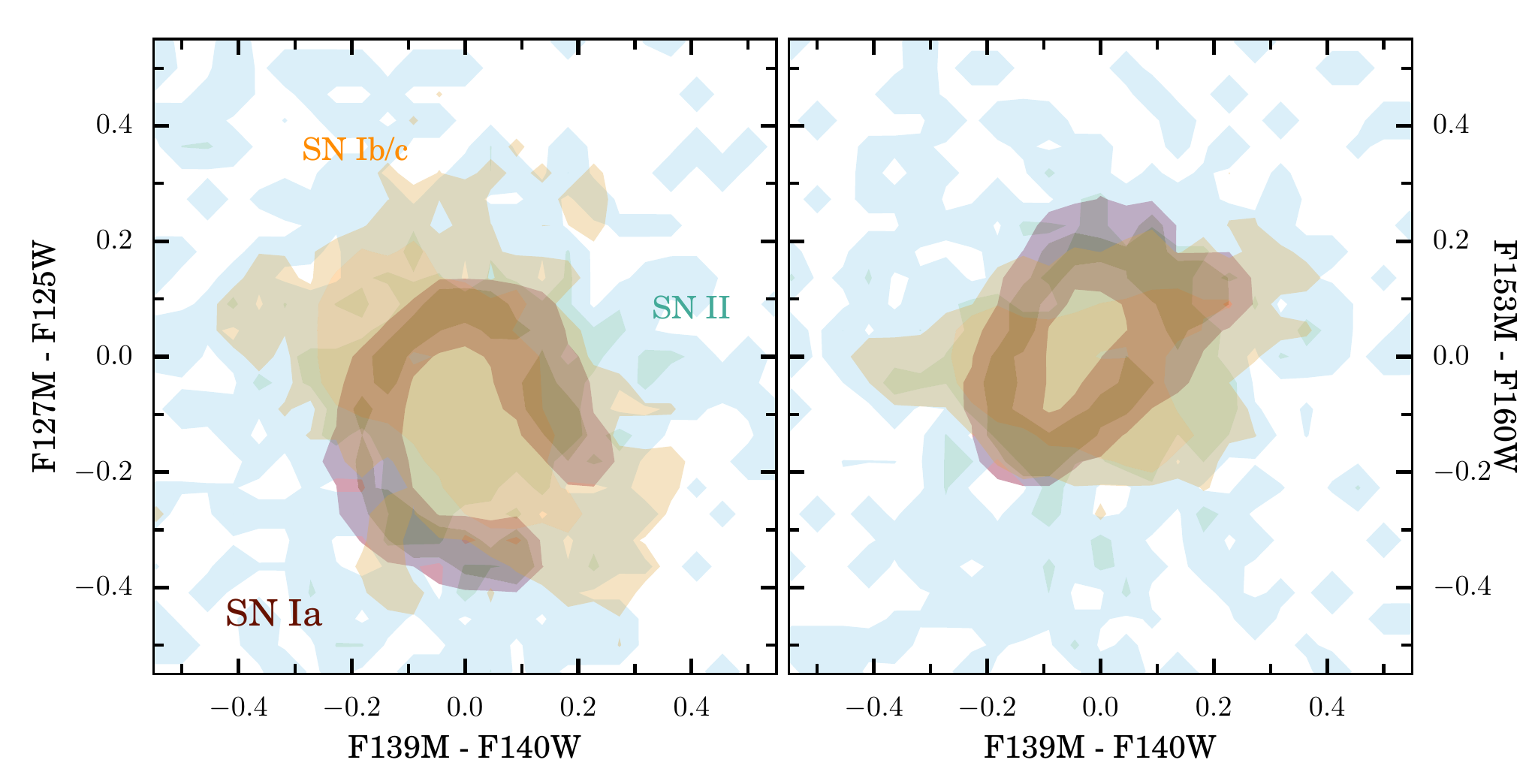}
\caption{  
Demonstration of the segregation of SN populations in {\it
medium$-$wide} pseudo-color space.  Contours show the colors measured
from 5000 simulated SN of each class, evenly sampling the redshift
range z=[1.8,2.2], with Monte Carlo sampling of light curve templates
and parameters -- including luminosity and host galaxy extinction --
following \R14.  Solid line contours enclose 68\% of the population
and filled contours enclose 95\% for each class, with red for \SNeIa,
green for Type II, and yellow for Type Ib/c. The separation of the
Type Ia contours from the \CCSN\ sub-types means that these
pseudo-color plots can be useful aids for classifying observed
high-$z$ SNe.
\label{fig:ClassifyDemo} }
\end{center}
\end{figure*}

The medium-band pseudo-colors are also useful for classification of
SNe at $z>1.5$.  The spectra of core collapse SNe (\CCSNe) do not have
the same semi-periodic absorption features as in \SNeIa.  \CCSN\ SEDs
are also much more heterogeneous, so the space that they populate in a
pseudo-color versus pseudo-color plot is not a circle, but an amorphous
cloud.  As shown in Figure~\ref{fig:ClassifyDemo}, the most common
sub-class of \CCSNe\ (Type II) is centered near the origin of the
pseudo-color space, so the location of a SN in this
space can be a useful indicator of its class.

\begin{figure*}
\begin{center}
\includegraphics[width=\textwidth]{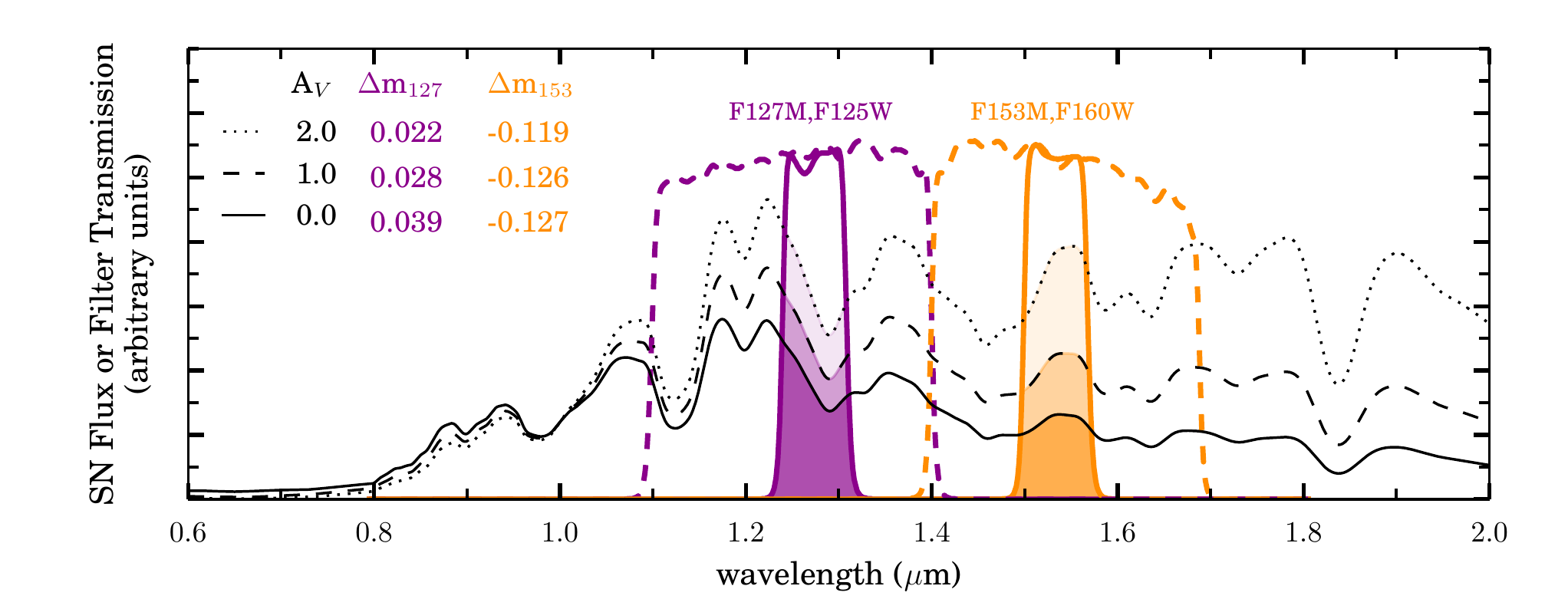}
\caption{
The medium-band pseudo-colors are largely insensitive to extinction.
Black curves show the SED of a Type Ia SN template at z=2.0 and peak
brightness \citep{Hsiao:2007}. Three curves show the effect of host
galaxy extinctions from \Av=0.0 to 2.0, using an extinction law
with \Rv=3.1 \citep{Cardelli:1989,ODonnell:1994}. The SEDs have been
normalized at 1 $\mu$m.  Although the overall SED is substantially
reddened, the pseudo-colors change by only $\sim$2\%, as given in the
legend in AB mag colors for F127M-F125W and F153M-F160W.
\label{fig:ExtinctionDemo} }
\end{center}
\end{figure*}

The use of medium-band pseudo-colors has several qualitative features
that make it attractive for SN classification.  First, by using
relative flux instead of absolute flux measures (\ie\ colors instead
of magnitudes), this method does not require any assumptions about the
underlying cosmology or luminosity function of the high-redshift SN
populations.  Second, the wide- and medium-band filters in each color
pair occupy the same portion of the SED, so this method is insensitive
to the overall color of the SED. Therefore, as shown in
Figure~\ref{fig:ExtinctionDemo}, the classification is largely
unaffected by SN host galaxy extinction: medium$-$wide pseudo-colors
are changed by less than 0.01 mag per 1 mag of extinction.

Medium-band imaging is also substantially more efficient than
spectroscopy.  As shown in \citet{Jones:2013}, observing a
typical \SNIa\ at $z>1.5$ with the WFC3-IR G141 grism requires at
least 10 HST orbits (25 ks) to get sufficient signal-to-noise (S/N)
for discriminating a \SNIa\ spectrum from that of a Type II SN. To
rule out possible mis-classification of Type Ib or Ic SNe as \SNIa\
requires considerably more: as many as $\sim$70 orbits at $z\sim2$.
HST grism observations are also often limited by systematic
uncertainties due to contamination from the SN host galaxy and other
nearby sources that overlap the SN spectrum. 
\changeG{By contrast, in this medium band imaging method we use spectral
elements that have $\sim$15\% better total throughput at any given
wavelength than the G141 grism.  }

With medium-band imaging we can also utilize existing wide-band
template images (already needed for measuring the wide-band light
curve) to subtract out contaminating background light very
cleanly. \change{That is, we get the advantages of host galaxy
subtraction without requiring an additional post-SN epoch of
medium-band imaging.  To do so, we make the justifiable assumption
that the host galaxy does not have any broad spectral features that
match the width of the medium band filters. With that assumption, we
can create a SN-free medium-band template image by simply scaling the
flux of the wide-band template using a ratio of the medium- and
wide-band filter transmissions.}

 In the following sections we will examine two examples from the
CANDELS SN program where medium-band imaging is employed at a total
cost of only 5-10 HST orbits ($\sim$10-20 ks) across all medium-band
filters.

\section{SN Discovery and Observations}\label{sec:Observations}

We now turn from the theory of this technique to the actual practice,
as applied to a pair of SNe discovered in the CANDELS multi-cycle
treasury program on HST.  The SN component of the CANDELS program was
operated jointly alongside the Cluster Lensing And Supernova Survey
with Hubble (CLASH, \citealt{Postman:2012}), and the two programs
together discovered and characterized $\sim$100 SNe over 3 years
(\citealt{Rodney:2012}, \citealt{Jones:2013}, \citetalias{Rodney:2014}, \citealt{Graur:2014}, \citealt{Patel:2014})
Follow-up observations with HST were executed through the
CANDELS/CLASH SN program (PI:Riess, HST Program ID:12099).  The SNe in
this work were first presented in \citetalias{Rodney:2014}, where we
measured the volumetric \SNIa\ rate as a function of redshift from the
full CANDELS SN sample.

SN \stone\ (Section~\ref{sec:Stone}) and \colfax\
(Section~\ref{sec:Colfax}) were discovered in IR imaging on the
CANDELS GOODS-S and GOODS-N fields, respectively.  In both cases these
SN were immediately identified as likely high-redshift sources based
on preliminary host galaxy photo-$z$ measurements and their observed
brightnesses.  These SNe were also initially classified as Type Ia
candidates, based on their rest-frame UV-optical color
deficit \citep[as in][]{Riess:2004a,Rodney:2012,Jones:2013}. No
definitive spectroscopic redshifts were available for either of these
objects at the time of discovery, and they were both quite faint, so
spectroscopic follow-up with the HST grisms was deemed too
risky. Instead, we applied the medium band imaging technique,
described in detail below.

Table \ref{tab:SNDiscovery} summarizes the discovery dates,
coordinates, and host galaxy redshifts of all three objects.
In \R14 we provided classifications and redshifts based only on the
broad-band SN light curves and host information. Here we will refine
both class and redshift with the addition of medium-band IR data and
improved photometry. We also extend the analysis to determine luminosity
distances for the two objects classified as Type Ia, in order to
compare them to the larger SN samples at $z<1$.

\subsection{Data Processing and Photometry}\label{sec:Photometry}

  \begin{deluxetable*}{llllll} 
\tablecolumns{6}
\tablecaption{SN Discovery Data\label{tab:SNDiscovery}}
\tablehead{ 
  \colhead{Supernova}
  & \colhead{SN Coordinates\tablenotemark{a}}
  & \colhead{Host Galaxy Coordinates\tablenotemark{b}}
  & \colhead{Discovery Date}
  & \colhead{Host Redshift}
  & \colhead{Host z Source\tablenotemark{c}}
}
\startdata
\stone      & 12:37:16.77 $+$62:16:41.4 & 12:37:16.59 $+$62:16:43.4 & UT 2013 Jun14.5 & $1.80\pm0.02$ & Spitzer IRS \\
\colfax     & 12:36:37.58 $+$62:18:33.0 & 12:36:37.51 $+$62:18:32.6 & UT 2012 May27.3 & $2.26\pm0.09$ & CANDELS phot-z\\
\bush\tablenotemark{d} & 03:32:42.78 $-$27:48:07.1 & 03:32:42.78 $-$27:48:07.1 & UT 2011 Aug05.8 & $1.15^{+0.19}_{-0.38}$  & CANDELS phot-z\\
\enddata
\tablenotetext{a}{R.A. and Decl. in J2000 ICRS coordinates.}
\tablenotetext{b}{The center of the SN host galaxy in J2000 ICRS coordinates. For \stone\ we give coordinates of the most likely host: stoneA.}
\tablenotetext{c}{Primary source for host galaxy redshift, independent of the SN photometry.  See text for details.}
\tablenotetext{d}{SN \bush\ is discussed in the appendix.}
\end{deluxetable*}

For each SN, we collected from the HST archive all available imaging
from the Wide Field Camera 3 (WFC3-IR and WFC3-UVIS) and the Advanced
Camera for Surveys Wide Field Camera (ACS-WFC).  These data were then
processed using a custom pipeline {\tt
sndrizpipe}\footnote{{\tt sndrizpipe} v1.2 \url{https://github.com/srodney/sndrizpipe}} that sorts the HST images into epochs, then
combines and registers them using the {\tt DrizzlePac} software from
the Space Telescope Science Institute (STScI) \citep{Fruchter:2010}.
For each filter, a template image was constructed from all epochs that
were at least 30 days before the SN appeared or 90 days after the last
3$\sigma$ detection.  The template images were directly subtracted
from every epoch containing SN light to get difference images.  It was
not necessary to convolve with the point spread function (PSF) for
optimal subtraction \citep{Alard:1998}, owing to the excellent
stability of the HST PSF.

\begin{deluxetable*}{rrrrrrrrr} 
\tablecolumns{9}
\tablecaption{\stone\ Photometry\label{tab:stonephot}}
\tablehead{ 
    \colhead{Obs. Date}
  & \colhead{Filter}
  & \colhead{Exp. Time}
  & \colhead{Flux}
  & \colhead{Flux Err}
  & \colhead{AB Mag\tablenotemark{a}}
  & \colhead{Mag Err}
  & \colhead{AB Zero Point} 
  & \colhead{$\Delta$ZP\tablenotemark{b}} \\
    \colhead{(MJD)}
  & \colhead{}
  & \colhead{(sec)}
  & \colhead{(counts/sec)}
  & \colhead{(counts/sec)}
  & \colhead{}
  & \colhead{}
  & \colhead{} 
  & \colhead{(Vega-AB)} 
}
\startdata
56403.4 & F350LP &  434 &    0.0236  &  0.1375  & $>$27.9 &  \nodata &  26.949 & -0.156\\
56457.5 & F350LP &  434 &    0.1084  &  0.1430  & $>$27.9 &  \nodata &  26.949 & -0.156\\
56490.6 & F350LP &  544 &    0.1271  &  0.1292  & $>$28.0 &  \nodata &  26.949 & -0.156\\
56513.4 & F350LP &  434 &   -0.0748  &  0.1548  & $>$27.8 &  \nodata &  26.949 & -0.156\\[1mm]
56406.7 &  F814W & 6708 &    0.0864  &  0.0675  &  28.606 &    0.849 &  25.947 & -0.424\\
56458.3 &  F814W & 2236 &    0.1512  &  0.0876  &  27.998 &    0.629 &  25.947 & -0.424\\[1mm]
56410.7 & F850LP & 1946 &   -0.1063  &  0.0917  & $>$26.3 &  \nodata &  24.857 & -0.519\\[1mm]
56403.4 &  F125W & 1156 &    0.2437  &  0.1155  &  27.763 &    0.515 &  26.230 & -0.901\\
56457.5 &  F125W & 1156 &    1.1137  &  0.1564  &  26.113 &    0.152 &  26.230 & -0.901\\
56467.9 &  F125W & 1256 &    2.1157  &  0.1614  &  25.416 &    0.083 &  26.230 & -0.901\\
56490.6 &  F125W & 1006 &    2.3250  &  0.1907  &  25.314 &    0.089 &  26.230 & -0.901\\
56513.4 &  F125W & 1156 &    1.2606  &  0.1651  &  25.979 &    0.142 &  26.230 & -0.901\\
56531.1 &  F125W &  503 &    0.4965  &  0.1907  &  26.990 &    0.417 &  26.230 & -0.901\\[1mm]
56474.4 &  F139M & 5023 &    0.4202  &  0.0791  &  25.420 &    0.204 &  24.479 & -1.076\\[1mm]
56474.5 &  F140W & 1356 &    2.7573  &  0.3110  &  25.351 &    0.122 &  26.452 & -1.076\\[1mm]
56474.9 &  F153M & 5023 &    0.4106  &  0.0621  &  25.429 &    0.164 &  24.463 & -1.254\\[1mm]
56403.4 &  F160W & 1206 &    0.1636  &  0.1493  &  27.911 &    0.991 &  25.946 & -1.251\\
56457.5 &  F160W & 1206 &    1.0590  &  0.1611  &  25.884 &    0.165 &  25.946 & -1.251\\
56467.9 &  F160W & 1206 &    1.2366  &  0.1543  &  25.715 &    0.136 &  25.946 & -1.251\\
56474.5 &  F160W & 1206 &    1.9397  &  0.1501  &  25.227 &    0.084 &  25.946 & -1.251\\
56490.6 &  F160W & 1006 &    1.5039  &  0.1728  &  25.503 &    0.125 &  25.946 & -1.251\\
56513.4 &  F160W & 1206 &    1.3238  &  0.1740  &  25.641 &    0.143 &  25.946 & -1.251\\
56531.1 &  F160W & 1609 &    0.9151  &  0.1628  &  26.042 &    0.193 &  25.946 & -1.251\\
56551.2 &  F160W & 2462 &    0.7125  &  0.1277  &  26.314 &    0.195 &  25.946 & -1.251\\
\enddata
\tablenotetext{a}{For \change{flux values of less than 1$\sigma$ significance} we report the magnitude as a 3$\sigma$ upper limit}
\tablenotetext{b}{Zero point difference: the magnitude shift for conversion from AB to Vega magnitude units.}
\end{deluxetable*}

\begin{deluxetable*}{rrrrrrrrr} 
\tablecolumns{9}
\tablecaption{\colfax\ Photometry\label{tab:colfaxphot}}
\tablehead{ 
    \colhead{Obs. Date}
  & \colhead{Filter}
  & \colhead{Exp. Time}
  & \colhead{Flux}
  & \colhead{Flux Err}
  & \colhead{AB Mag\tablenotemark{a}}
  & \colhead{Mag Err}
  & \colhead{AB Zero Point} 
  & \colhead{$\Delta$ZP\tablenotemark{b}} \\
    \colhead{(MJD)}
  & \colhead{}
  & \colhead{(sec)}
  & \colhead{(counts/sec)}
  & \colhead{(counts/sec)}
  & \colhead{}
  & \colhead{}
  & \colhead{} 
  & \colhead{(Vega-AB)} 
}
\startdata
56018.7  &  F350LP  &    434 &      0.2079  & 0.1657  & 28.654  &    0.865  &  26.949 & -0.156\\
56074.3  &  F350LP  &    868 &     -0.2289  & 0.1649  & $>$27.7 &  \nodata  &  26.949 & -0.156\\
56129.4  &  F350LP  &    434 &     -0.3159  & 0.2412  & $>$27.3 &  \nodata  &  26.949 & -0.156\\
56145.5  &  F350LP  &    400 &     -0.0398  & 0.1953  & $>$27.5 &  \nodata  &  26.949 & -0.156\\[1mm]
56070.0  &   F606W  &    982 &     -0.0505  & 0.1123  & $>$27.7 &  \nodata  &  26.493 & -0.086\\[1mm]
56016.3  &   F814W  &   2536 &      0.1885  & 0.1006  & 27.759  &    0.580   &  25.947 & -0.424\\
56074.6  &   F814W  &   3535 &      0.1318  & 0.0952  & 28.147  &    0.784   &  25.947 & -0.424\\
56116.5  &   F814W  &   7072 &     -0.0889  & 0.0746  & $>$27.6 &  \nodata   &  25.947 & -0.424\\
56128.4  &   F814W  &   6708 &     -0.0630  & 0.0834  & $>$27.5 &  \nodata   &  25.947 & -0.424\\
56180.9  &   F814W  &   2236 &     -0.0433  & 0.1186  & $>$27.1 &  \nodata   &  25.947 & -0.424\\
56200.2  &   F814W  &   2396 &      0.0735  & 0.0479  & 28.781  &    0.708   &  25.947 & -0.424\\[1mm]
56152.1  &  F850LP  &   1442 &     -0.0647  & 0.0746  & $>$26.5 &  \nodata   &  24.857 & -0.519\\[1mm]
56018.6  &   F125W  &   1156 &     -0.0049  & 0.0703  & $>$27.9 &  \nodata   &  26.230 & -0.901\\
56074.3  &   F125W  &   1156 &      1.2037  & 0.1440  & 26.029  &    0.130   &  26.230 & -0.901\\
56084.7  &   F125W  &   1206 &      1.1206  & 0.1487  & 26.106  &    0.144   &  26.230 & -0.901\\
56103.2  &   F125W  &   1206 &      0.8749  & 0.1514  & 26.375  &    0.188   &  26.230 & -0.901\\
56129.5  &   F125W  &   2309 &      0.2662  & 0.0990  & 27.667  &    0.404   &  26.230 & -0.901\\
56145.6  &   F125W  &   2562 &      0.2219  & 0.1033  & 27.864  &    0.505   &  26.230 & -0.901\\
56183.1  &   F125W  &   1156 &     -0.0035  & 0.0708  & $>$27.9 &  \nodata   &  26.230 & -0.901\\
56242.0  &   F125W  &   1156 &      0.0000  & 0.0763  & $>$27.8 &  \nodata   &  26.230 & -0.901\\
56297.7  &   F125W  &   1156 &      0.0140  & 0.0742  & $>$27.9 &  \nodata   &  26.230 & -0.901\\[1mm]
56084.8  &   F127M  &   7485 &      0.3853  & 0.0447  & 25.677  &    0.126   &  24.641 & -0.961\\[1mm]
56082.8  &   F139M  &   8291 &      0.3472  & 0.0527  & 25.628  &    0.165   &  24.479 & -1.079\\[1mm]
56083.1  &   F140W  &   1706 &      1.9370  & 0.1847  & 25.734  &    0.103   &  26.452 & -1.076\\
56091.5  &   F140W  &    506 &      1.8524  & 0.2847  & 25.783  &    0.167   &  26.452 & -1.076\\
56104.2  &   F140W  &    506 &      1.3515  & 0.2988  & 26.125  &    0.240   &  26.452 & -1.076\\[1mm]
56085.7  &   F153M  &   7485 &      0.2942  & 0.0543  & 25.791  &    0.201   &  24.463 & -1.254\\[1mm]
56018.6  &   F160W  &   1206 &     -0.0116  & 0.1823  & $>$26.6 &  \nodata   &  25.946 & -1.251\\
56074.3  &   F160W  &   1206 &      1.3706  & 0.1727  & 25.604  &    0.137   &  25.946 & -1.251\\
56084.7  &   F160W  &   1206 &      1.0138  & 0.1547  & 25.931  &    0.166   &  25.946 & -1.251\\
56103.2  &   F160W  &   1256 &      0.9008  & 0.1550  & 26.059  &    0.187   &  25.946 & -1.251\\
56129.4  &   F160W  &   2762 &      0.4049  & 0.1231  & 26.928  &    0.330   &  25.946 & -1.251\\
56145.5  &   F160W  &   2009 &      0.5715  & 0.1500  & 26.553  &    0.285   &  25.946 & -1.251\\
56183.1  &   F160W  &   1206 &      0.2813  & 0.1489  & 27.323  &    0.575   &  25.946 & -1.251\\
56242.0  &   F160W  &   1206 &      0.0664  & 0.2891  & $>$26.1 &  \nodata   &  25.946 & -1.251\\
56297.7  &   F160W  &   1206 &     -0.0234  & 0.2204  & $>$26.4 &  \nodata   &  25.946 & -1.251\\
\enddata
\tablenotetext{a}{For \change{flux values of less than 1$\sigma$ significance} we report the magnitude as a 3$\sigma$ upper limit}
\tablenotetext{b}{Zero point difference: the magnitude shift for conversion from AB to Vega magnitude units.}
\end{deluxetable*}

Fluxes and AB magnitudes for  \stone\ and \colfax\ are
given in Tables \ref{tab:stonephot}
and \ref{tab:colfaxphot}, respectively.  In the WFC3-IR difference
images we measured the SN flux using PSF-fitting photometry, with a
PSF model constructed from the G2V standard star P330E. Flux
uncertainties were estimated by separately planting 10,000 synthetic SNe
into the difference image and fitting a Gaussian to the distribution
of recovered fluxes.

 For both of these SNe we have no detections above 3$\sigma$ in
the ACS-WFC or WFC3-UVIS imaging.  In those bluer bands (corresponding
to the rest-frame UV for these SNe) we used the {\tt hstphot}
package\footnote{\url{https://github.com/srodney/hstphot}; based on a
translation of the IDL AstroLib photometry
routines \citep{Landsman:1993} into Python.} to 
collect aperture photometry in a 0.4\arcsec aperture, with
zero points and aperture corrections as published
by STScI.\footnote{For
WFC3: \url{http://www.stsci.edu/hst/wfc3/phot_zp_lbn}; for
ACS: \url{http://www.stsci.edu/hst/acs/analysis/zeropoints}}

\subsection{Galaxy Photometry, Host Association\\ and SED Fitting}\label{sec:HostMethods}

\begin{small}
  \begin{deluxetable*}{crrrccrrl} 
\tablecolumns{9}
\tablecaption{Kron Ellipse Parameters and Properties of Host Galaxy and Lens Candidates\label{tab:GalaxySEDFits}}
\tablehead{ 
    \colhead{ID} &
  $r_{\rm A}$ (\arcsec) & $r_{\rm B}$  (\arcsec) & $\theta$ (deg) & $R_{\rm SN}$
  & \colhead{spec-z}
  & \colhead{photo-z\,\tablenotemark{a}}
  & \colhead{$\log_{10}$(M/\Msun)\,\tablenotemark{b}}
  & \colhead{spec-z Ref.} 
}
\startdata
colfaxA & 0.34 &  0.25 &   0.0 &   1.6 &  \nodata              &   2.26 $^{+0.09}_{-0.09}$ &  9.98 $^{+0.10}_{-0.06}$ & \nodata \\[0.5em]
colfaxB & 0.64 &  0.46 &  89.4 &   4.4 &  0.9466 $\pm$ 0.0002  &   0.95 $^{+0.14}_{-0.04}$ &  9.87 $^{+0.17}_{-0.14}$ & \citealt{Cowie:2004} \\[0.5em]
stoneA  & 0.60 &  0.26 &  52.1 &   3.9 &  1.80 $\pm$ 0.02      &   1.70 $^{+0.23}_{-0.13}$ & 11.37 $^{+0.31}_{-0.12}$ & \citealt{MurphyEJ:2009}\\[0.5em]
stoneB  & 0.59 &  0.41 &  76.8 &   5.0 &  \nodata              &   1.55 $^{+0.66}_{-0.22}$ & 10.13 $^{+0.29}_{-0.25}$ & \nodata  \\[0.5em]
stoneC  & 0.66 &  0.31 &  51.1 &   5.5 &  0.5572 $\pm$ 0.0002  &   0.65 $^{+0.05}_{-0.07}$ &  9.23 $^{+0.20}_{-0.24}$ & \citealt{Wirth:2004}\\[0.5em]
bushA   & 0.15 &  0.13 &   1.1 &   0.0 &  \nodata              &   1.15 $^{+0.19}_{-0.38}$ &  7.60 $^{+0.20}_{-0.19}$ & \nodata \\[0.5em]
\enddata
\tablenotetext{a}{Photometric redshift and 68\% confidence region.}
\tablenotetext{b}{$\log_{10}$ of the total stellar mass of the galaxy and 68\% confidence region, with redshift fixed to the spec-z when available.}
\end{deluxetable*}
\end{small}

To evaluate galaxies in the vicinity of our SNe as candidate hosts and
potential gravitational lenses, we use a photometry catalog including
data from all available optical and IR bands. All photometry was
collected using SExtractor \citep{Bertin:1996} for source detection
and TFIT \citep{Laidler:2007} for matched-aperture photometry, as
detailed in \citet{Guo:2013} and \citet{Fontana:2014}. For CANDELS
data where the SN was detected (i.e. the WFC3 IR bands) we measured
photometry from the SN-free template images instead of the full-depth
CANDELS mosaics.

To determine which of the nearby galaxies (if any) could plausibly be
associated with each SN, we follow the procedure
of \citet{Sullivan:2006a}, which is designed to infer the most likely
SN host galaxy based on angular separation and galaxy size.  Using
SExtractor on the F160W images, we define a \citet{Kron:1980} ellipse
centered at position $x_{gal},y_{gal}$, with semimajor axis $r_A$,
semiminor axis $r_B$ and position angle $\theta$.  The separation from
the galaxy to any given position ($x,y$) is quantified by the
elliptical radius, $R$, defined as

\begin{equation}  \label{eqn:R}
   R^2 = C_{xx}x_r^2 + C_{yy}y_r^2 + C_{xy}x_r y_r,
\end{equation}

\noindent where $x_r=x-x_{gal}$,  $y_r=y-y_{gal}$,
$C_{xx}=\cos^2(\theta)/r_A^2 + \sin^2(\theta)/r_B^2$,
$C_{yy}=\sin^2(\theta)/r_A^2 + \cos^2(\theta)/r_B^2$, and
$C_{xy}=2\cos(\theta)\sin(\theta)(1/r_A^2 - 1/r_B^2)$. An ellipse with
radius $R=2.5$ will typically contain $\sim$90\%\ of the object's
light \citep{Infante:1987,Graham:2005}. For consideration as a
possible SN host galaxy, \citet{Sullivan:2006a} used the requirement
that the SN position must lie within an ellipse with $R=5$, and found
that $\sim$93\% of SNe in the Supernova Legacy Survey (SNLS) had at
least one detected host candidate meeting that criterion.  Ellipse
parameters for the nearest galaxies to each SN are reported in
Table~\ref{tab:GalaxySEDFits}.  Note that this approach does not make
use of any redshift information about the SN or the nearby galaxies.
In the following sections we will also bring to bear the available
spectroscopic and photometric redshift evidence.

To determine the photo-$z$ and estimate physical parameters, we fit
each galaxy's SED using the LePhare code for SED matching via $\chi^2$
minimization \citep{Arnouts:1999,Ilbert:2006}. We employ a template
library based on the PICKLES stellar spectra \citep{Pickles:1998}, the
LePhare quasar templates, and the BC03 models for galaxy
SEDs \citep{Bruzual:2003}.  A Chabrier initial mass function is
assumed \citep{Chabrier:2003}.  Further details of the SED-fitting
process are given in \citet{Hemmati:2014}. \changeG{Note, however,
that in \citet{Hemmati:2014} each galaxy is subdivided to kpc-scale
resolution. For this work we use the integrated light of the entire
galaxy.}  Table~\ref{tab:GalaxySEDFits} reports the results from this
SED fitting, giving the photometric redshift, best-fit galaxy type,
and an estimate of the total stellar mass.  We find the photometric
redshifts are in very good agreement with the few available
spectroscopic redshifts.  In those cases where a spectroscopic
redshift exists, the redshift is fixed at that value for determination
of the best-fit galaxy type and stellar mass.

\section{\stone\ : SN Stone}\label{sec:Stone}

\begin{figure}
\begin{center}
\includegraphics[width=\columnwidth]{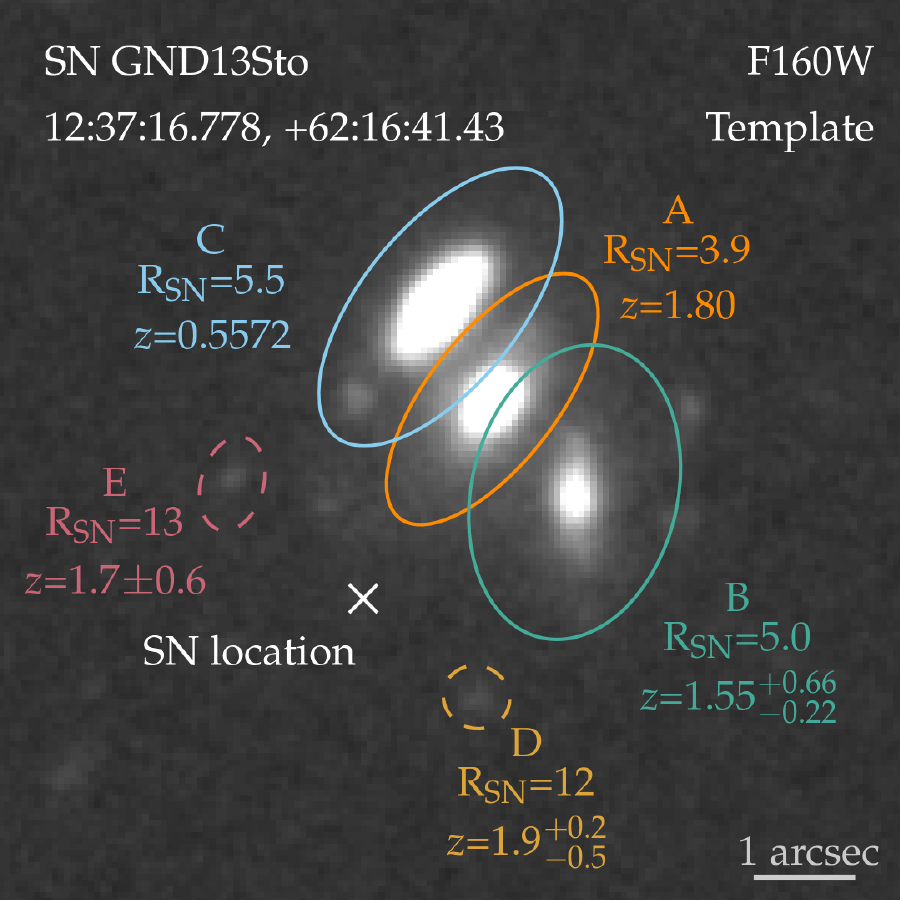}
\caption{  
The host environment of SN \stone.  The template F160W image shown is
a combination of all exposures from before MJD=56400 or after
MJD=56600. Ellipses are drawn at $R=2.5$ for the five nearest host
galaxy candidates, using parameters measured with SExtractor in the
F160W band (see Equation~\ref{eqn:R}).  Each host galaxy candidate is
labeled with the $R$ value of an ellipse that passes through the
position of the SN, and the best available redshift constraint. As
detailed in the text, galaxy A is the most probable host, while
galaxies B, D and E all have broad photo-$z$ distributions consistent
with the redshift of galaxy A.  Some or all of these four galaxies may be
members of a small group hosting SN \stone.
\label{fig:stonehost} }
\end{center}
\end{figure}

\begin{figure}
\begin{center}
\includegraphics[width=\columnwidth]{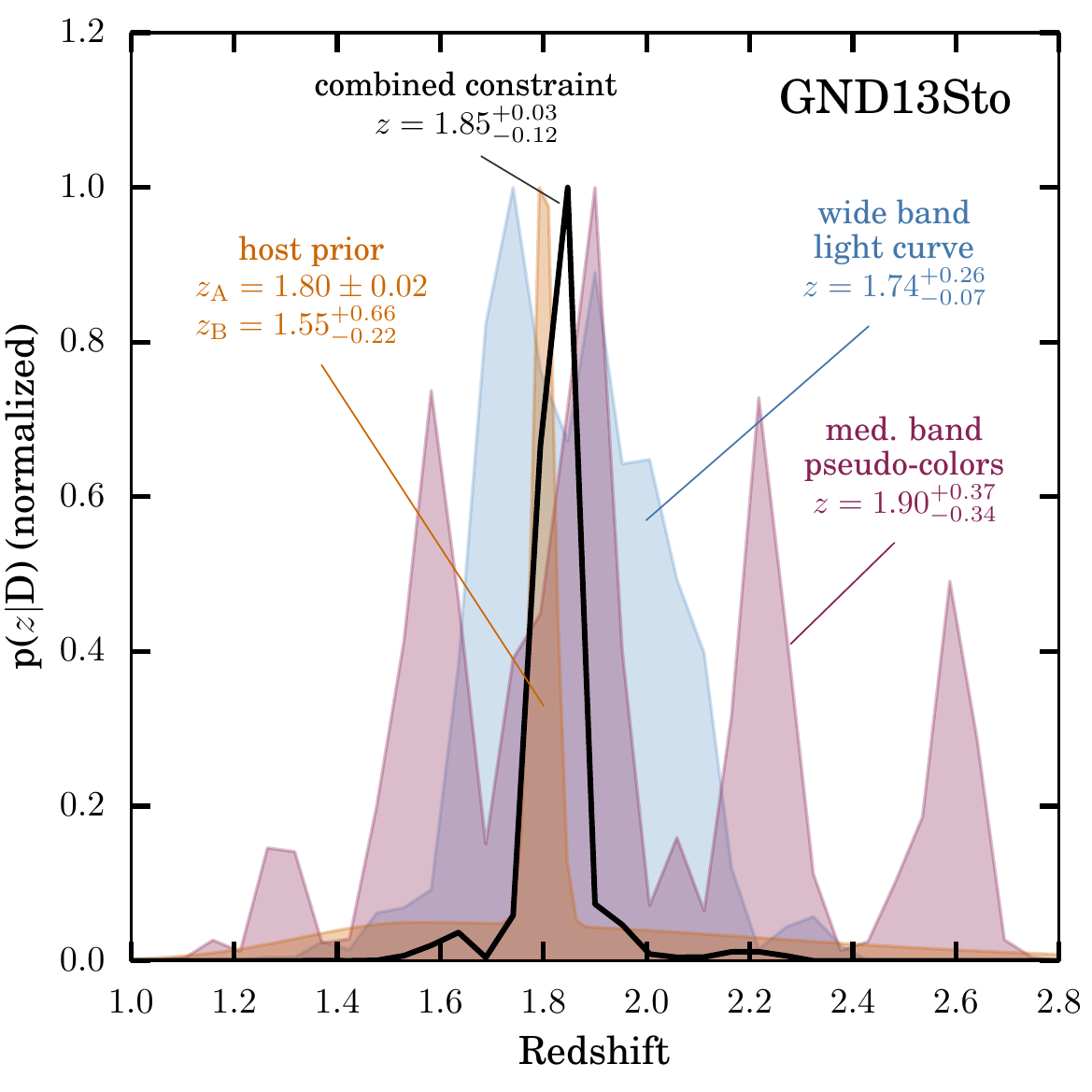}
\caption{  
Redshift constraints on \stone, assuming it is a Type Ia SN.  The
independent constraint from each of 3 different sources and the
combined constraint are shown as probability distribution functions
over redshift, all normalized to a peak value of 1.0.
The \change{composite redshift prior from host galaxy candidates A and
B} is given as a filled orange curve. The filled blue curve shows the
marginal posterior probability from light curve fitting with only the
wide-band data and no host galaxy redshift prior.  The
filled red curve shows the constraint from the single epoch of
medium band data, along with the corresponding wide-band
observations in the same epoch, but no host galaxy redshift
prior. Finally, the solid black curve shows the composite redshift
constraint derived from light curve fitting with all of the above
information.  }
\label{fig:stonepostz}
\end{center}
\end{figure}

\begin{figure*}
\begin{center}
\includegraphics[width=\textwidth]{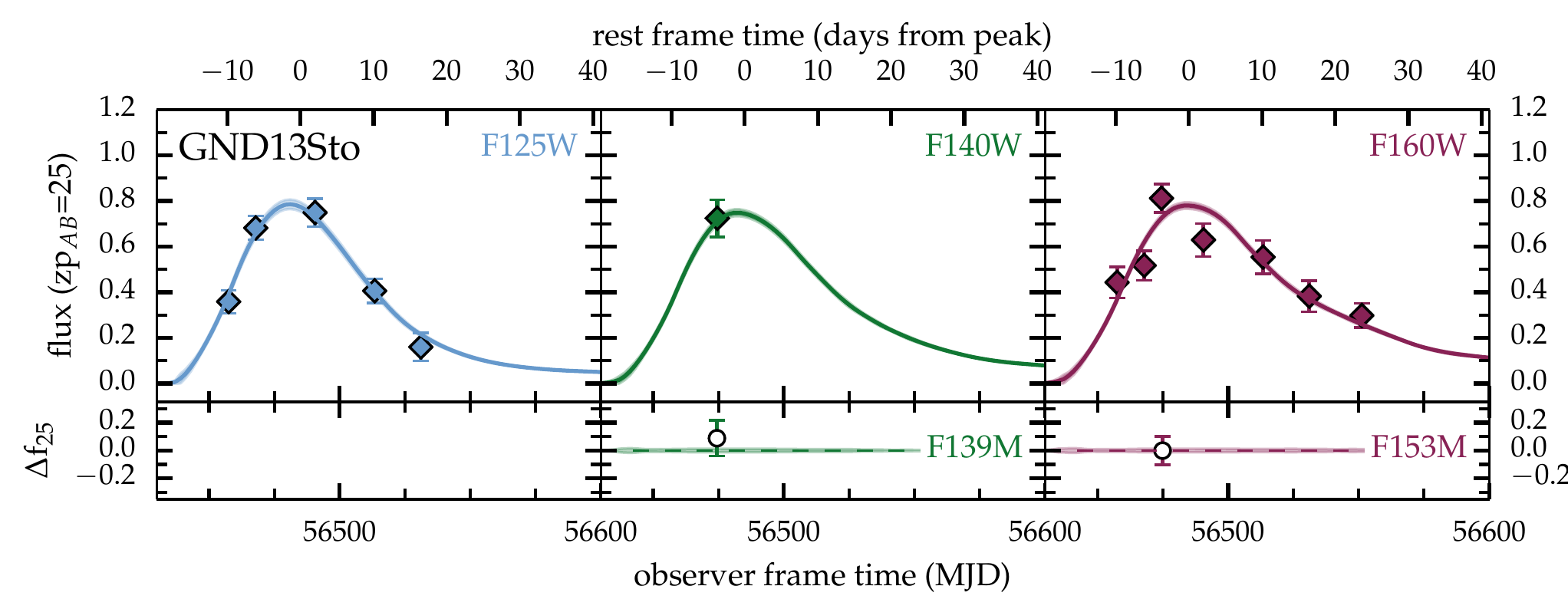}
\caption{  
Infrared light curve of \stone, showing flux as a function of time.
Top panels show observations in HST wide-bands, in flux units
normalized to an AB magnitude zero point of 25.  Lower panels show
{\it residual} fluxes for the corresponding HST medium-bands, after
subtracting off the predicted flux from the best-fit model.  The top
axis marks rest-frame time at redshift $z=1.8$.  relative to the time
of B band peak brightness.
\label{fig:stonelc} }
\end{center}
\end{figure*}

\begin{figure*}
\begin{center}
\includegraphics[width=\textwidth]{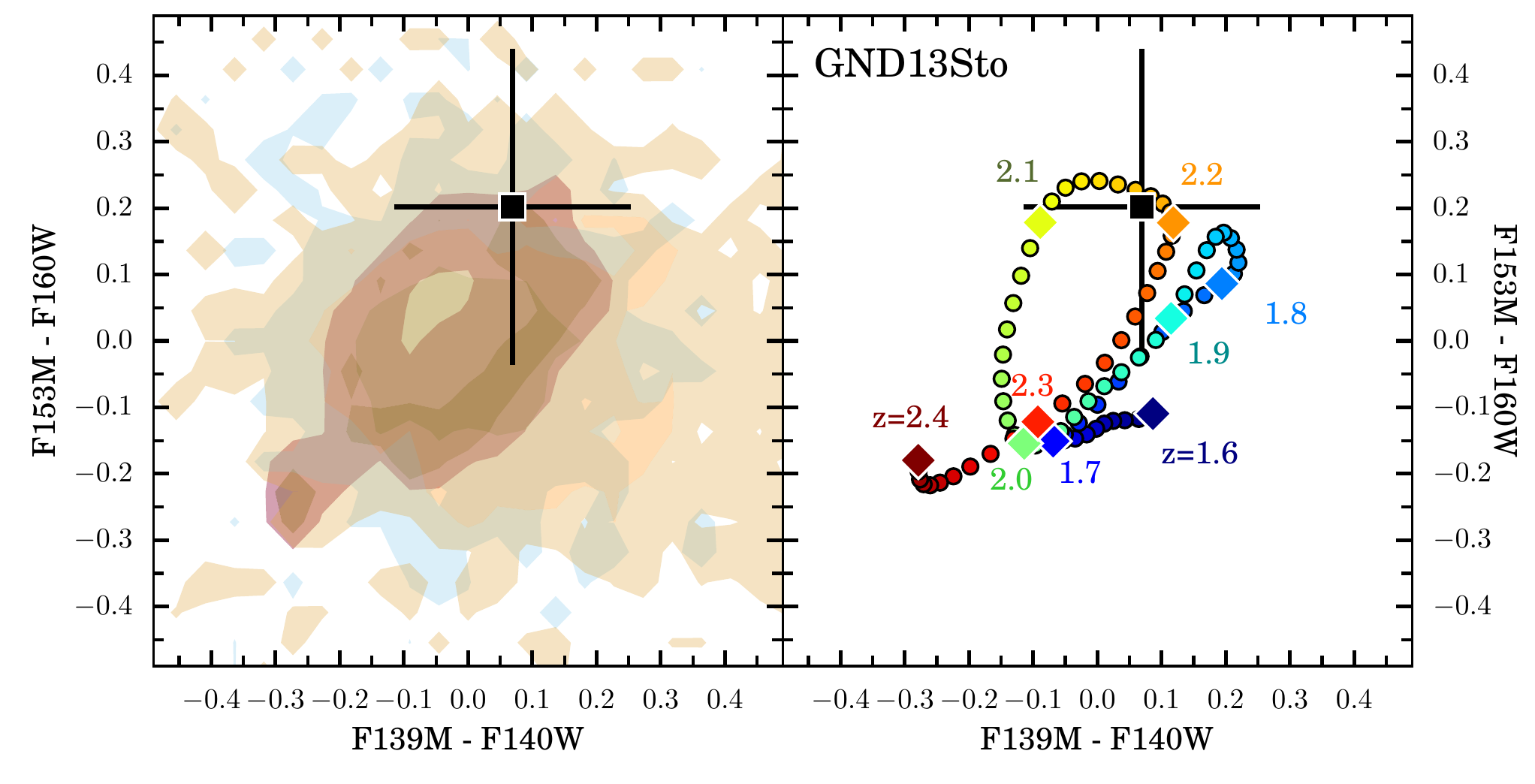}
\caption{  
Medium-band pseudo-color circle diagrams for \stone, as in
Figures~\ref{fig:RedshiftDemo} and \ref{fig:ClassifyDemo}.  In both
panels, the observed colors for \stone\ are indicated by the black
square with error bars.  ({\it Left}) Classification circle diagram:  contours enclose 68\% and 95\%
of the population for each class, with red for Type Ia, green for Type
II and gold for Type Ib/c.  The simulated SNe have redshifts that
evenly sample the redshift range [1.6,2.4].  ({\it Right}) Redshift circle diagram : the line of
colored points traces the pseudo-colors of a \SNIa\ from z=1.6 (blue)
to z=2.4 (red), with diamond symbols marking each increment of 0.1 in
redshift.  }
\label{fig:StoneCircle}
\end{center}
\end{figure*}

\subsection{\stone\ Host Galaxy}\label{sec:StoneHost}

The first of our two high-$z$ \SNIa\ candidates is \stone,
\footnote{Nicknamed ``SN Stone,'' after Thomas Stone, a signatory of
the United States Declaration of Independence as a delegate from
Maryland}
discovered in exposures taken on 14 June 2013 in the GOODS-N field.
There are 22 galaxies detected in the CANDELS F160W imaging that lie
within 10\arcsec\ of the SN position, but none of these are coincident
with the SN.  

Using the $R$ parameter (Equation~\ref{eqn:R}) as an indication of the
likelihood of physical association, we reduce the list of possible
hosts to the 3 galaxies highlighted in Figure~\ref{fig:stonehost},
which contain the SN position within an ellipse having $R<10$.  Of
these, candidate A is the most likely, with $R=3.9$, though host
candidates B and C are close behind, with $R=5.0$ and 5.5,
respectively.
\citet{MurphyEJ:2009} measured the redshift of galaxy A as  
$z=1.82\pm0.02$, based on the IR spectrum from the InfraRed
Spectrograph (IRS) on the \Spitzer\ space telescope, and it was 
subsequently updated to $z=1.80\pm0.02$ in \citet{Kirkpatrick:2012}.
\changeG{ This IRS spectrum
was also revisited by \citet{Hernan-Caballero:2012}, who used a
maximum-likelihood spectral template fitting approach that confirmed
the solution around
$z=1.8$. However, \citeauthor{Hernan-Caballero:2012} erroneously
associated the IRS spectrum with SDSS J123716.59+621643.9 (our galaxy C) and thus incorrectly identified it as an outlier.}

Host candidate B is separated from the SN by $R_{\rm SN}=5.0$, and has
a photometric redshift of $z=1.55^{+0.66}_{-0.22}$. Though quite
broad, this photo-$z$ overlaps the spectroscopic redshift of galaxy A,
$z=1.8$.  There are two other galaxies within 10\arcsec\ of the SN (D
and E in Figure~\ref{fig:stonehost}) that also have very broad
photometric redshifts consistent with galaxy A: $z=1.9^{+0.2}_{-0.5}$
for galaxy D and $z=1.7\pm0.6$ for E.  These two low surface
brightness galaxies are very unlikely to be directly associated
with \stone, as they are separated from the SN by $R_{\rm SN}>10$.
Nevertheless, the similarity of these photo-$z$ estimates suggests that
galaxies A and B may be two dominant members of a small galaxy group
containing D and E.  Given the weak redshift constraints, this is
very speculative, but if true then it is possible that SN \stone\ is
either bound to an undetected member of that group, or belongs to the
intracluster light.

The only remaining galaxy near enough to be related to the SN is host
candidate C.  This is the galaxy SDSS J123716.59+621643.9, which has a
spectroscopic redshift of $z=0.5572$ from an optical
spectrum \citep{Wirth:2004}.  We identify galaxy C as a foreground
object because the SN photometry is inconsistent with this redshift:
no normal SN light curve template at $z=0.56$ can match the observed
photometry with a $\chi^2$ per degree of freedom ($\chi^2_\nu$) better
than 3.3.

\change{
In summary, we find that \stone\ is most likely associated with either
galaxy A or galaxy B.  If the former, then the SN is at redshift
$z=1.8$, and if the latter then the redshift is weakly constrained by
the photo-z to $z=1.55^{+0.66}_{-0.22}$.  After classifying the SN in
Section \ref{sec:StoneClassification} we will revisit this redshift
constraint in Section \ref{sec:StoneClassification}, bringing to bear
all available information from the SN light curve, and also
considering the possibility of an unseen host galaxy. 
}

\subsection{\stone\ Classification}\label{sec:StoneClassification}

In the left panel of Figure~\ref{fig:StoneCircle} we consider the
classification information provided by the medium-band circle diagram
(developed in Section~\ref{sec:Method} and
Figure~\ref{fig:ClassifyDemo}).  The observed medium-band
pseudo-colors of SN \stone\ are modestly offset from the origin in this
pseudo-color space, locating it in the circular region where a Type Ia SN is expected to
lie.  The uncertainties are large, so although
Figure~\ref{fig:StoneCircle} is suggestive of a Type Ia
classification, by itself this evidence from medium-band pseudo-colors
is far from definitive.  Recall, however, that this information from
the medium-band imaging is particularly valuable because it is
insensitive to reddening and it is indicative of the presence or
absence of specific absorption features in the SN's SED.

For a more complete classification picture, we next classify \stone\
using a Bayesian photometric classification algorithm, as
in \citetalias{Rodney:2014}.  This classifier \change{uses the {\tt
sncosmo} software package\footnote{{{\tt sncosmo} v1.1, source code
at \url{https://github.com/sncosmo}, documentation
at \url{https://sncosmo.readthedocs.org/en/v1.0.x/}}} to generate
simulated SN light curves from spectrophotometric templates.} We rely
on the SALT2 model \citep{Guy:2010} to represent normal Type Ia SN,
and use 42 spectrophotometric templates from the {\tt SNANA} template
library\footnote{{\tt SNANA}: SuperNova ANAlysis
package; \url{http://das.sdss2.org/ge/sample/sdsssn/SNANA-PUBLIC/}} \citep{Kessler:2009a}
to represent CC SNe (26 Type II and 16 Type Ib/c).  \change{Although
this template library is at present the best available collection of
spectrophotometric models for CC SNe, it is an incomplete
sampling of this heterogeneous population. It is possible to
incorporate some knowledge about this incompleteness into the
photometric classification framework \citep[e.g.][]{Rodney:2009}, but
for this work we make the simplifying assumption that the library
provides a nearly complete and evenly distributed sampling of CC SN
properties.}

For redshifts $z>1.5$, the available templates (particularly those
for \CCSNe) do not extend far enough into the rest-frame UV to allow
for a comparison to observations in ACS-WFC or WFC3-UVIS filters
without extrapolating the templates.  Thus, in this analysis we
restrict our photometric classifications and light curve fitting to
the available WFC3-IR observations, which correspond to rest-frame
optical bands. As noted above, we have only upper limits for ACS and
UVIS fluxes for both of these SN, so this stricture does not
substantially diminish the available information.

\change{
We have made one important change in the classification algorithm
relative to previous work, which is to eschew a luminosity prior.  To
get photometric classifications of the complete CANDELS/CLASH sample
in \citetalias{Rodney:2014} and \citet{Graur:2014} we included a soft
prior on the peak absolute magnitude as a function of redshift for
each SN sub-class. The prior was based on the observed luminosity
functions of these SN types in the local universe combined with an
assumed cosmological model.  This was necessary for the calculation of
volumetric SN rates, as a large fraction of the sample has very
limited photometric data (in some cases just 1-2 epochs with 2 or 3
bandpasses).  Although in principle one could employ a weak prior
without introducing a significant bias, in this work we have enough
data such that we can get robust classifications without the
luminosity/cosmology prior. This ensures that these SNe can be used
for cosmology without any concern of a bias introduced at the
classification stage.
}

In the case of \stone, we first classify the object using \change{both
wide- and medium-band photometry} but including no redshift prior, and
find that it is strongly classified as a Type Ia SN. The net
classification probability for all CC SN sub-types is $<1\%$, and no
CC template can match the photometric observations with $\chi^2_\nu$
better than 2.8.  This agrees with our prior analysis
in \citetalias{Rodney:2014}, where we found that \stone\ is robustly
classified as a Type Ia SN based on a Bayesian photometric
classification using the wide-band light curves.

Our photometric classification apparatus can also accommodate a prior
that infers the SN class based on the color and morphology of the host
galaxy \citep{Foley:2013a}.  In this case, using either of the
plausible host candidates, galaxies A and B, we still find
that \stone\ is classified as a \SNIa\ with $>98\%$ confidence.

\subsection{\stone\ Redshift}\label{sec:StoneRedshift}

Due to the ambiguity of the host association, it is particularly
valuable to evaluate the redshift constraints on \stone\ independent
of any host.  As a first step, in the right panel of
Figure~\ref{fig:StoneCircle} we consider the redshift information
provided by the medium-band pseudo-color circle diagram (see
Figure~\ref{fig:RedshiftDemo}).  Working now under the assumption that
SN \stone\ is a Type Ia SN, the location in the upper right corner of
this pseudo-color space is most consistent with a redshift near
$z=1.85$ or near $z=2.2$. Redshifts close to 1.7, 2.0, and 2.4 are
disfavored by these data.  Due to the low S/N of the medium-band
observations for this SN, this diagram by itself is not definitive for
redshift determination.  However, as we will see, the addition of
medium band observations to a complete wide-band light curve does
provide a valuable improvement in the redshift precision, and
supports an association with host galaxy A at $z=1.8$.

In Figure~\ref{fig:stonepostz} we build up a composite redshift
constraint piece by piece, showing how the posterior probability
distribution over redshift is separately restricted by the host
galaxy, the wide-band light curve, and the colors from a single epoch
that includes medium-band observations.  First, the wide band light
curve alone provides a fairly weak constraint, $z=1.74^{+0.26}_{-0.07}$.  
Second, from the single epoch of medium-band pseudo-colors collected
near peak brightness we derive a posterior probability distribution
that is periodic in redshift.  With only a single epoch, the {\it
shape} of the light curve is unconstrained (i.e. any change in the
value of $x_1$ could be reproduced by coordinated changes in $x_0$ and
$t_{\rm pk}$).  To address this, we limit the range of $t_{pk}$ to
$\pm3$ days around the value inferred from the wide-band light curve
fit -- effectively adding in additional information by fixing the
phase of the light curve at the time of the medium-band observations.
Formally, we derive the redshift and uncertainty as 
\change{$z=1.90^{+0.37}_{-0.34}$},
but Figure~\ref{fig:stonepostz} shows that we actually have 4 peaks,
at $z\sim1.5$, 1.9, 2.2, and 2.6.  This periodicity arises from
spectral features moving in and out of the medium bands as the SN Ia
template redshift changes, illustrated in
Figure~\ref{fig:RedshiftDemo} and previously observed in \S09.  By
combining the wide band light curve and the medium band epoch
together, this redshift degeneracy is broken.

\change{
The final piece of redshift information we would like to include is a
host galaxy redshift prior.  Given that the host galaxy association is
ambiguous, we use a composite redshift constraint that is the sum of
the redshift probability distributions from host galaxy candidates A
and B.  In effect this allows an equal probability that the SN belongs
to either galaxy, resulting in a sharp peak at z=1.8 (from host
candidate A) and broad tails on either side (from host candidate B),
as shown in Figure~\ref{fig:stonepostz}. Including this as a redshift
prior, and utilizing all available SN photometry, we get a final
redshift constraint of $z_{\rm A+B}=1.85^{+0.03}_{-0.12}$.}  To test
how sensitive this redshift is to the host galaxy prior, we have also
computed the redshift of \stone\ using only the photometric redshift
of host candidate B ($1.55^{+0.66}_{-0.22}$).  This results in a
posterior probability distribution that is peaked very close to the
spectroscopic redshift of host candidate A: 
$z_{\rm B}=1.85^{+0.10}_{-0.19}$.

\change{A final possibility to entertain is that the true host galaxy of
SN \stone\ is not detected in the CANDELS imaging
data.  \citet{Sullivan:2006a} found that 7\%\ of the SNe in the SNLS
program did not have a detected galaxy within $R_{\rm SN}=5$. Setting
aside SN \stone\, the equivalent ``hostless'' fraction is only 3\% in
the CANDELS/CLASH programs \citep{Graur:2014,Rodney:2014}, owing
largely to the advantage of very deep HST imaging in these survey
fields.  Nevertheless, to consider the possibility of an unseen host, we
repeat the redshift test using a flat prior, and find $z_{\rm no
host}=1.95^{+0.05}_{-0.26}$.  Thus, even if the \stone\ host is
undetected, the SN photometry alone requires the redshift to be 
fairly close to the spectroscopic redshift of host candidate A
(1.80$\pm$0.02).  }

\change{
We therefore conclude that the most likely scenario has SN \stone\ as
a Type Ia SN at $z=1.80\pm0.02$.  It may be bound to 
galaxy A, or associated with some possibly unseen component of
a group containing that galaxy.}  
At redshift $z=1.80$ the best-fit
SALT2 light curve parameters and uncertainties are
$t_{pk}=56482.1\pm1.2$, $x_1=-0.47\pm0.68$, $c=-0.02\pm0.07$, and
$m_{B}=26.14\pm0.07$.  Figure~\ref{fig:stonelc} shows this maximum
likelihood Type Ia model, which has $\chi^2=10.9$ for 12 degrees of
freedom.

\subsection{\stone\ Lensing}\label{sec:StoneLensing}

The foreground galaxy C at $z=0.5572$ is separated from the SN by
2\farcs9, making it a possible source of gravitational lensing for
SN \stone. Our procedure for estimating the lensing magnification
follows \citet{Jones:2013}.  

To define the lensing potential, we start with the stellar mass of the
galaxy derived from SED fitting (Table~\ref{tab:GalaxySEDFits}) and
use the broken power law relation of \citet{Yang:2012} to convert this
to the total mass of the dark matter halo.  We then model the stellar
component of the galaxy with a S\'ersic profile \citep{Sersic:1963} with
parameters derived using GALFIT \citep{Peng:2002}.  Next, we assume a
Navarro-Frenk-White (NFW) density profile \citep{Navarro:1997} for the
lensing potential, and use GRAVLENS \citep{Keeton:2001} to generate
10,000 Monte Carlo realizations in which we vary the stellar mass of
the lensing galaxy using the limits from SED fitting as reported in
Table~\ref{tab:GalaxySEDFits}, and the redshift of SN~\stone\ from
section~\ref{sec:StoneRedshift}.  Our Monte Carlo simulation also
varies the mass-concentration relation and the stellar-to-halo mass
ratio within their observed ranges \citep{Guo:2011,Yang:2012}.  From
this we infer a median magnification and 68\%\ confidence limits of
$1.02^{+0.02}_{-0.01}$ from the stoneC galaxy at the position of
SN~\stone, indicating that the SN is not substantially biased by
gravitational lensing.  We include no correction for the insignificant
magnification derived here, but see
Section~\ref{sec:GravitationalLensing} for discussion of lensing
effects on a full high-$z$ SN sample.

\section{\colfax : SN Colfax}\label{sec:Colfax}

\begin{figure*}
\begin{center}
\includegraphics[width=0.37\textwidth]{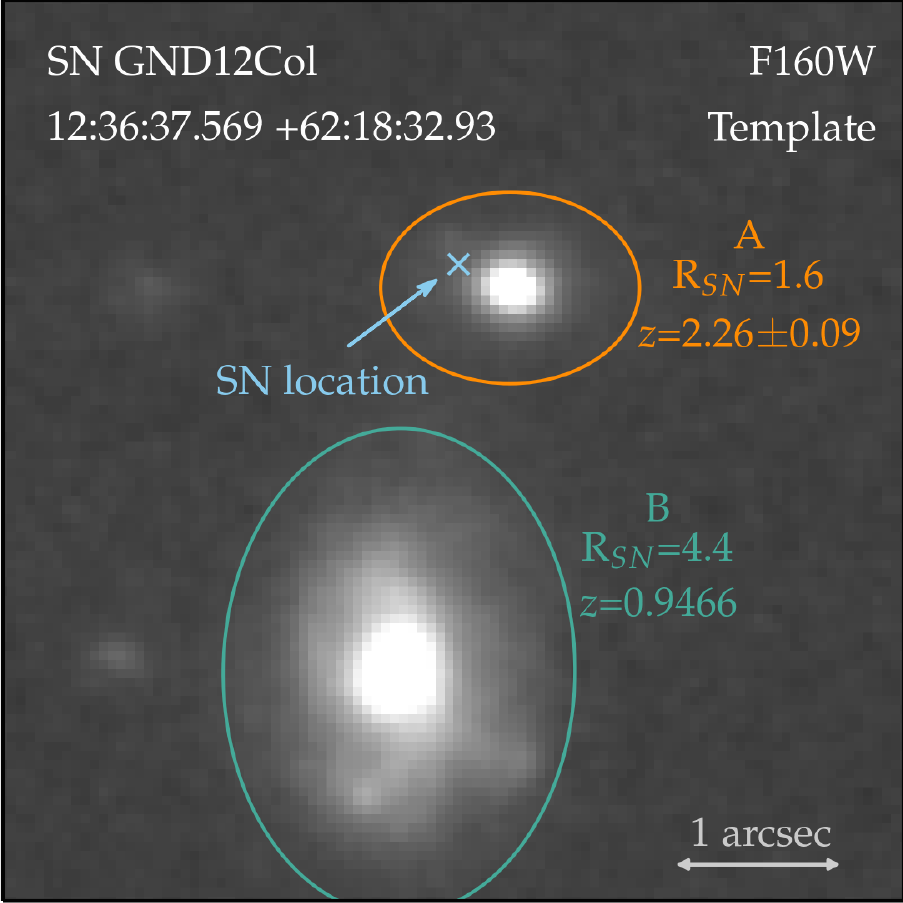}
\includegraphics[width=0.62\textwidth]{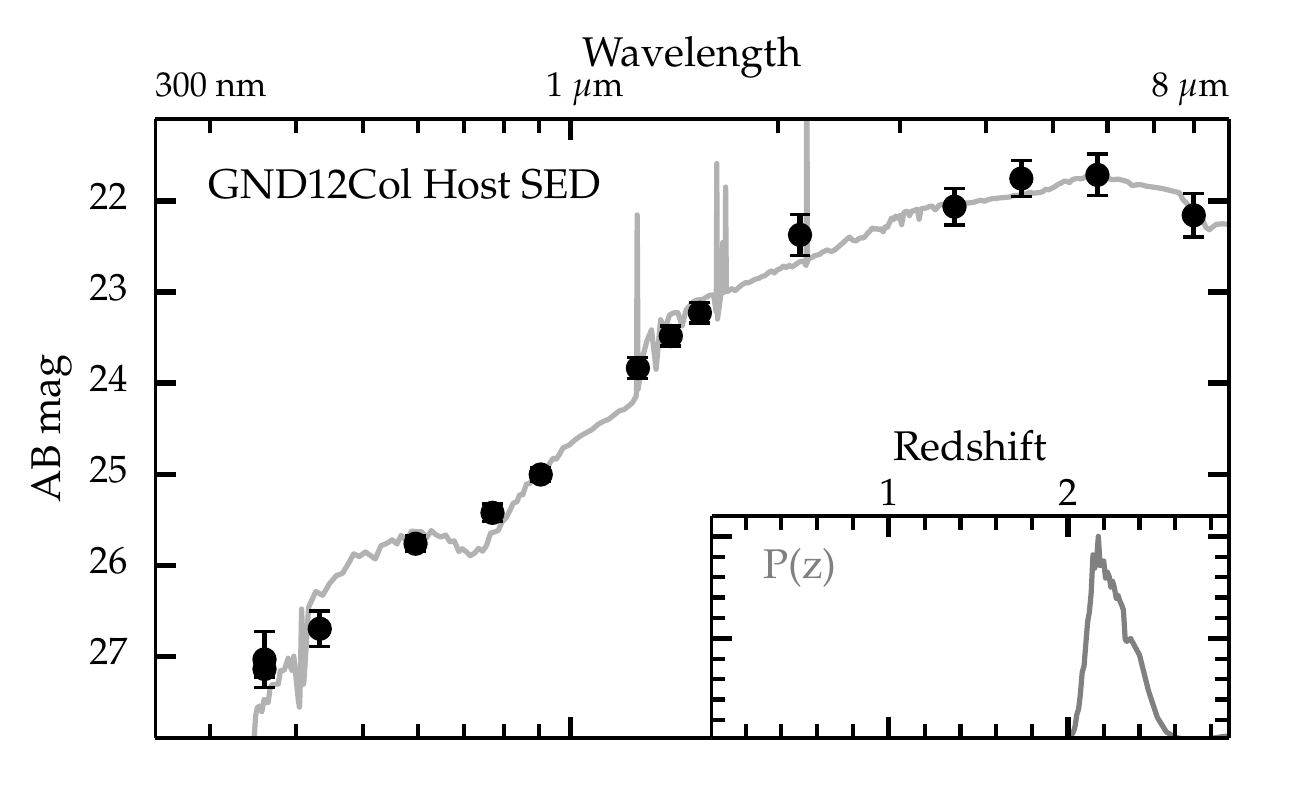}
\caption{  
The host galaxy of \colfax.  {\it Left:} Template F160W image
combining all exposures from before MJD=56000 or after
MJD=56300. Ellipses are drawn at $R=2.5$, using parameters
measured using SExtractor in the F160W band (see
Equation~\ref{eqn:R}).  Galaxy B is a foreground object.  {\it Right:}
Points show the observed photometry of the host, galaxy A, and the
curve shows the best-fitting SED model.  The inset at lower right
shows the redshift probability distribution function \change{derived
using the LePhare SED fitting code}, centered at $z=2.26$, with a
68\%\ confidence range of [2.17,2.35].
\label{fig:colfaxhost} }
\end{center}
\end{figure*}

\begin{figure*}
\begin{center}
\includegraphics[width=\textwidth]{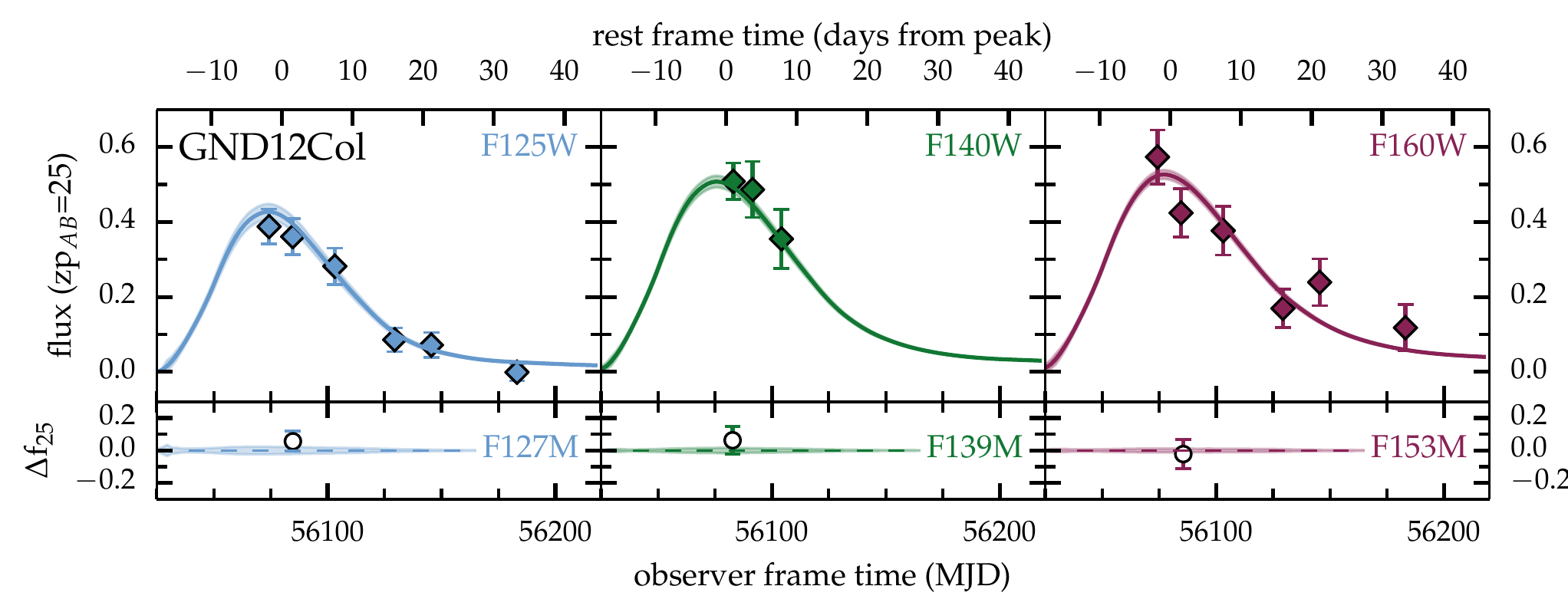}
\caption{  
Infrared light curve of \colfax, showing flux as a function of
time. As in Figure~\ref{fig:stonelc} fluxes are 
normalized to zero point 25 and lower panels show
residual fluxes for medium-band observations. The top
axis marks rest-frame time at redshift $z=2.26$.
\label{fig:colfaxlc} }
\end{center}
\end{figure*}

\begin{figure*}
\begin{center}
\includegraphics[width=\textwidth]{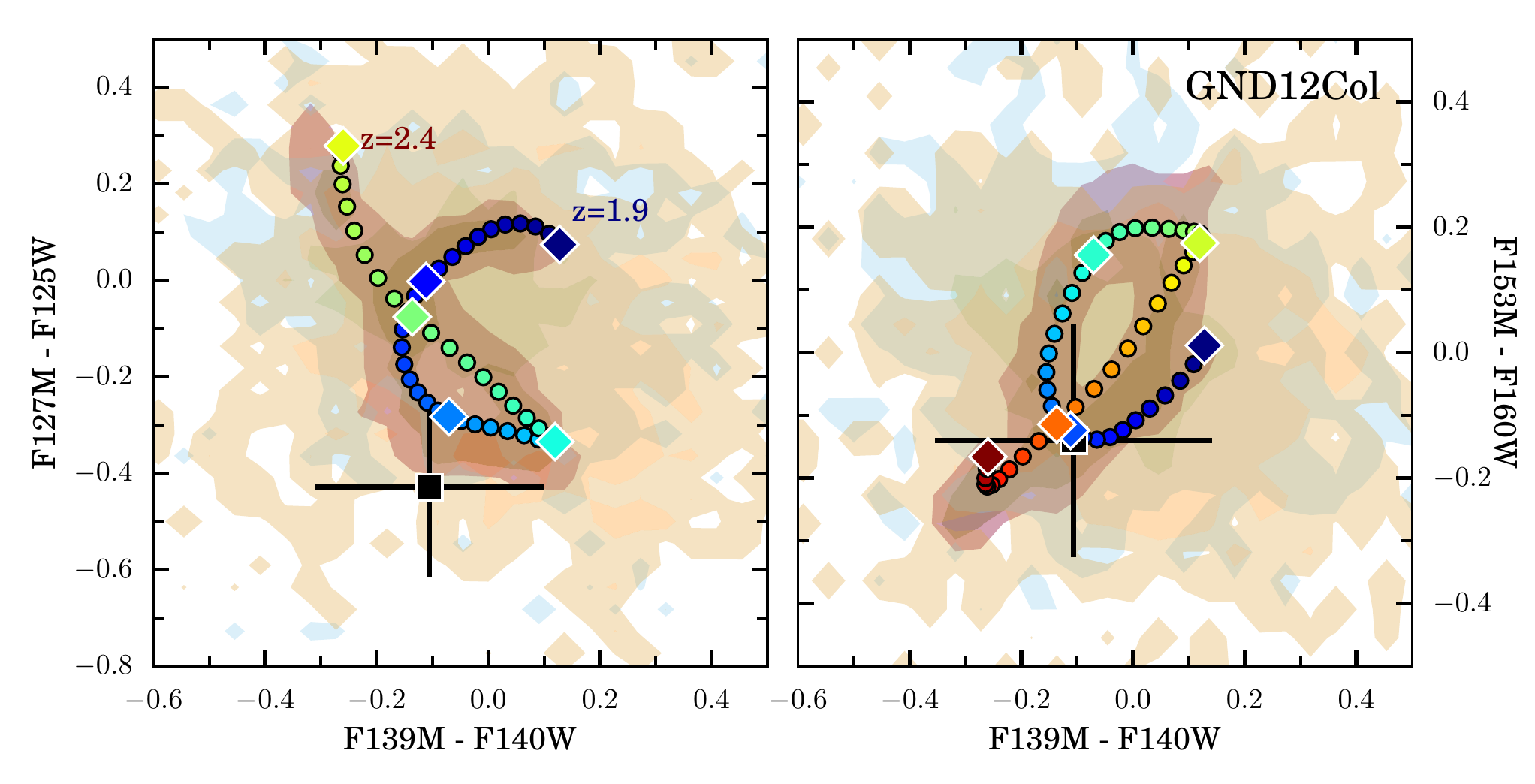}
\caption{  
Medium-band pseudo-color diagrams for \colfax, as in
Figures~\ref{fig:RedshiftDemo} and \ref{fig:ClassifyDemo}. Contours
enclose 68\% and 95\% of the population for each class, with red for
Type Ia, green for Type II and gold for Type Ib/c.  The simulated SN
have redshifts that evenly sample the redshift range [1.9,2.4].  The
line of colored points traces the pseudo-color of a \SNIa\ from z=1.9 (blue)
to z=2.4 (red), with diamond symbols marking each increment of 0.1 in
redshift.  Observed colors for \colfax\ are indicated by the
black square with error bars.}
\label{fig:ColfaxCircle}
\end{center}
\end{figure*}

\begin{figure}
\begin{center}
\includegraphics[width=\columnwidth]{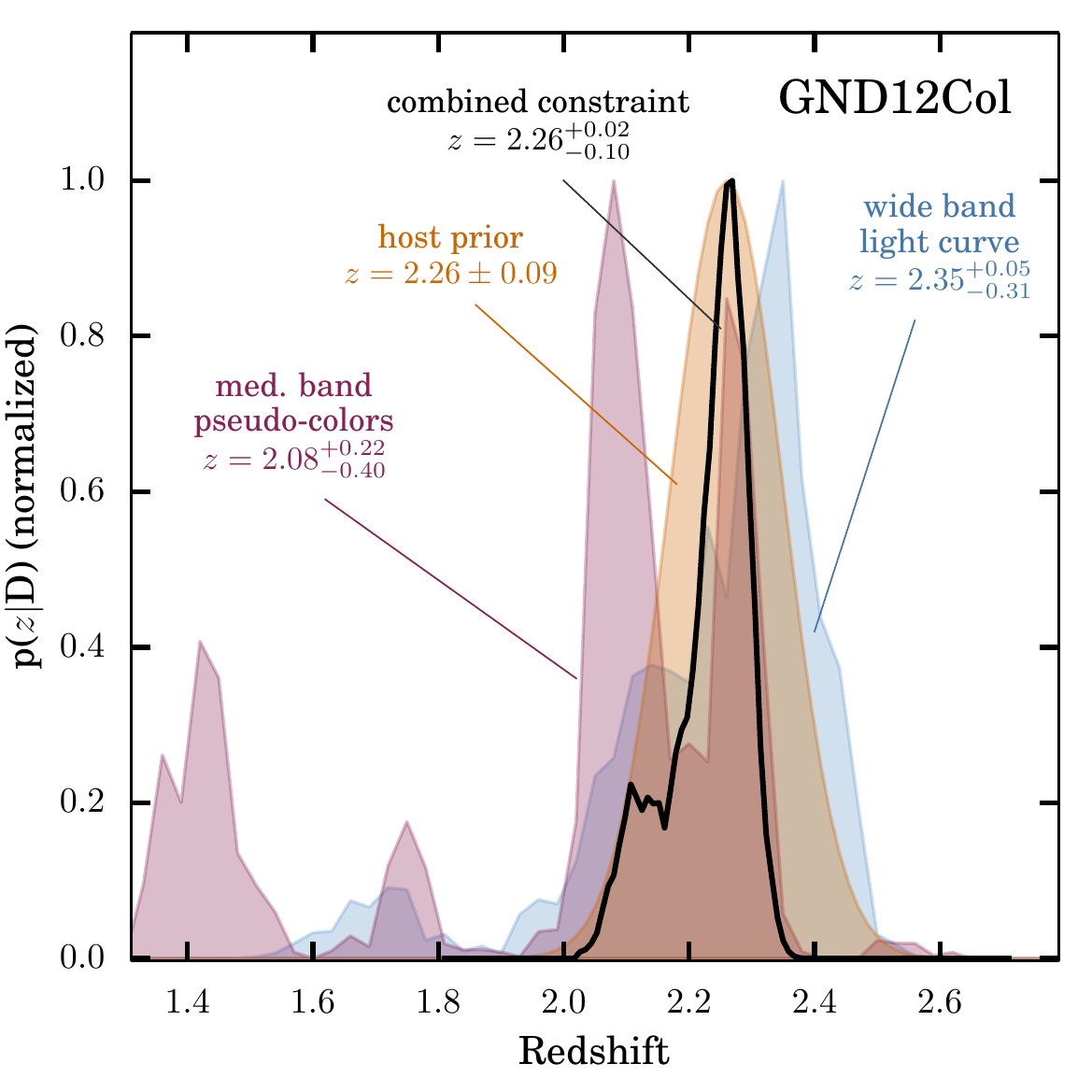}
\caption{  
Redshift constraints on \colfax.  As in Figure~\ref{fig:stonepostz},
the filled orange curve shows the host galaxy redshift prior (from
galaxy colfaxA), filled curves in blue and red show the constraints
from the wide-band light curve and the medium-band colors (without
including a redshift prior), and the solid black curve shows the
composite constraint from all available data.  All PDFs are normalized
to a peak value of 1.0.  }
\label{fig:colfaxpostz}
\end{center}
\end{figure}

\subsection{\colfax\ Host Galaxy}\label{sec:ColfaxHost}

Our second high-$z$ \SNIa\ candidate is \colfax, 
\footnote{Nicknamed ``SN Colfax,'' after Schuyler Colfax, 17th Vice President of the United States.}
found in the CANDELS GOODS-N field on 27 May 2012.  The nearest
galaxy to the SN position, labeled host candidate A in
Figure~\ref{fig:colfaxhost}, has a relatively tight photo-z at
$z=2.35$, with 68.3\% confidence interval [2.24,2.43].

\citet{Dahlen:2013} recently demonstrated that even well-constrained 
photometric redshift estimates from any single SED fitting algorithm
can be systematically biased and underestimate the redshift
uncertainties, though a combination of photo-$z$ estimates from
multiple codes can reduce these concerns. To address this, we also fit
the SED of the \colfax\ host candidate A using the {\it BPZ} code, a
Bayesian photometric redshift estimator \citep{Benitez:2000}.  We
derive a redshift probability distribution function, \change{$P(z)$,}
centered at $z=2.17$ with 68.3\% confidence region [2.10,2.23], in
good agreement with the photo-$z$ derived above.  We define the
composite photo-$z$ for host candidate A as a Gaussian centered at
the mean of the two, with a 68.3\% confidence region encompassing the
\change{mean of the two photo-$z$ distributions:} $z=2.26\pm0.09$.

A second possible host galaxy -- candidate B -- is the large face-on
spiral to the southeast, which has a spectroscopic redshift of
$z=0.9466\pm0.0002$ \citep{Cowie:2004}.  To determine whether galaxy A
or B is the more likely host, we first evaluate the SN-galaxy
separation independently of redshift using the $R$ parameter of
Equation~\ref{eqn:R}.  For host candidate A we find $R=1.61$, and for
host candidate B we have $R=4.4$, supporting candidate A as the more
likely host.  Including the available redshift information strengthens
this conclusion.  \colfax\ is 2.8\arcsec\ from the nucleus of host
candidate B, which is 23 kpc at the redshift of that galaxy (assuming
a flat $\Lambda$CDM cosmology with \Ho=70, \Om=0.3).  The SN is only
0.44\arcsec\ (3.3 kpc at $z\approx2.2$) from the center of host
candidate A.  Based on this evidence, we assign host
candidate A as the host of SN \colfax, lending the SN a strong
redshift prior at $z=2.26\pm0.09$.

\subsection{\colfax\ Classification and Redshift}\label{sec:ColfaxClassificationRedshift}

\citetalias{Rodney:2014} classified \colfax\ as a Type Ia SN, and we
find that this classification holds up with improved photometry and is
supported by the addition of medium-band observations.  The formal
Type Ia classification probability is \change{effectively unity}, with the
combined probability for all common \CCSN\ sub-classes limited to
$<10^{-6}$.  
Figure~\ref{fig:colfaxlc} shows the maximum
likelihood \SNIa\ model at $z=2.26$, which has $\chi^2$=19.8 for 21
degrees of freedom.  The best-fit SALT2 light curve parameters and
uncertainties are $t_{pk}=56077.4\pm5.6$, $x_1=0.15\pm1.06$, color
$c=0.04\pm0.13$, and peak apparent magnitude $m_{\rm B}=26.80\pm0.07$.

To visualize how the medium-band observations lead to an improvement
of the \colfax\ classification and redshift, we show in
Figure~\ref{fig:ColfaxCircle} the position of the SN in \change{a pair
of 2D projections} of the pseudo-color space. The \CCSN\ population is
expected to be spread out across the entire pseudo-color space, but in
both of these slices we find \colfax\ is conspicuously close to the
narrower region occupied by \SNeIa.  Note also that the SN position in
both projections of pseudo-color-color space is consistent with the
middle of the redshift distribution, i.e. $z\sim2.2$.

Figure~\ref{fig:colfaxpostz} shows how the posterior probability
distribution over redshift for \colfax\ is separately constrained by
the host galaxy prior, the wide-band light curve, and the colors from
a single epoch that includes medium-band observations.  Each of these
components alone provides a weak constraint, but all three are
consistent and with the combination we achieve a
precision of $\sigma_{z}=^{+0.02}_{-0.10}$, on par with the best available
photometric redshifts from well-sampled galaxy SEDs at similar
redshift \citep{Dahlen:2013}.

\subsection{\colfax\ Lensing}\label{sec:ColfaxLensing}

As in Section~\ref{sec:StoneLensing} we evaluate the possibility of
gravitational lensing from the foreground galaxy B.  This galaxy has a
projected separation of 2\farcs8, and a total stellar mass of
$\sim$10$^{10}$ \Msun.  As with \stone, we adopt a S\'ersic profile for
the stellar component and assume an NFW dark matter halo, \change{with
mass concentrations and stellar-to-halo mass ratios
following \citet{Guo:2011} and \citet{Yang:2012}.}  From 10,000 Monte
Carlo realizations of the lensing potential we determine a
magnification factor of $\mu=1.04^{+0.03}_{-0.02}$.  Once again we do
not include a correction to the SN magnification for this mild
magnification, but examine global lensing effects in
Section~\ref{sec:GravitationalLensing}.

\changeG{
\section{Evaluating the Medium-band\\ Pseudo-spectroscopy Technique}\label{sec:MedbandDiscussion}
}

\changeG{
For the two high-$z$ \SNeIa\ discussed in this work, one may well ask
whether the medium-band IR observations have provided significant
value for the SN classification and redshift determination.  After
all, both of these objects were classified as Type Ia SN with high
confidence in \R14, \change{and were assigned redshifts from their host
galaxies,} all without the benefit of the detailed medium-band analysis
employed here.
}

\change{Before scrutinizing the value of the medium band imaging in the
following sections, let us first note that relative to photometric
data the medium-band data are less susceptible to systematic biases.}
This is because (a) the medium-band pseudo-colors are insensitive to
reddening (Figure~\ref{fig:ExtinctionDemo}) and (b) the redshift
information in the medium-band data comes from individual spectral
absorption features (Figure~\ref{fig:RedshiftDemo}), not from the
overall shape of the SED or the light curve.

\change{
\subsection{Classification}\label{sec:ClassificationDiscussion}
}

\changeG{
\change{
A key practical benefit of the medium-band approach is that it can be
used for efficient direction of scarce follow-up resources toward
high-value targets. } \changeG{ When a high-$z$ \SNIa\ candidate is
initially discovered, a single epoch of medium-band imaging can
substantially clarify the classification picture while the SN is still
close to peak brightness -- as long as the redshift and time of peak
are fairly well constrained.  Thus, one can get a strong
classification assessment for a relatively low cost and then evaluate
whether the candidate warrants an investment of additional follow-up
time to collect the full wide-band light curves. } \change{
Importantly, the medium band pseudo-colors are insensitive to dust
(cf. Figure~\ref{fig:ExtinctionDemo}) and unaffected by the underlying
cosmology.  Thus, medium-band observations are similar to spectroscopy
in that they can be used to select a Type Ia sample for follow-up
without needing to invoke any priors on luminosity or color, which may
carry a bias with redshift. The medium band classifications are
therefore more robust than purely broad-band photometric methods, but
more efficient than spectroscopy.}
}

\begin{figure}
\begin{center}
\includegraphics[width=\columnwidth]{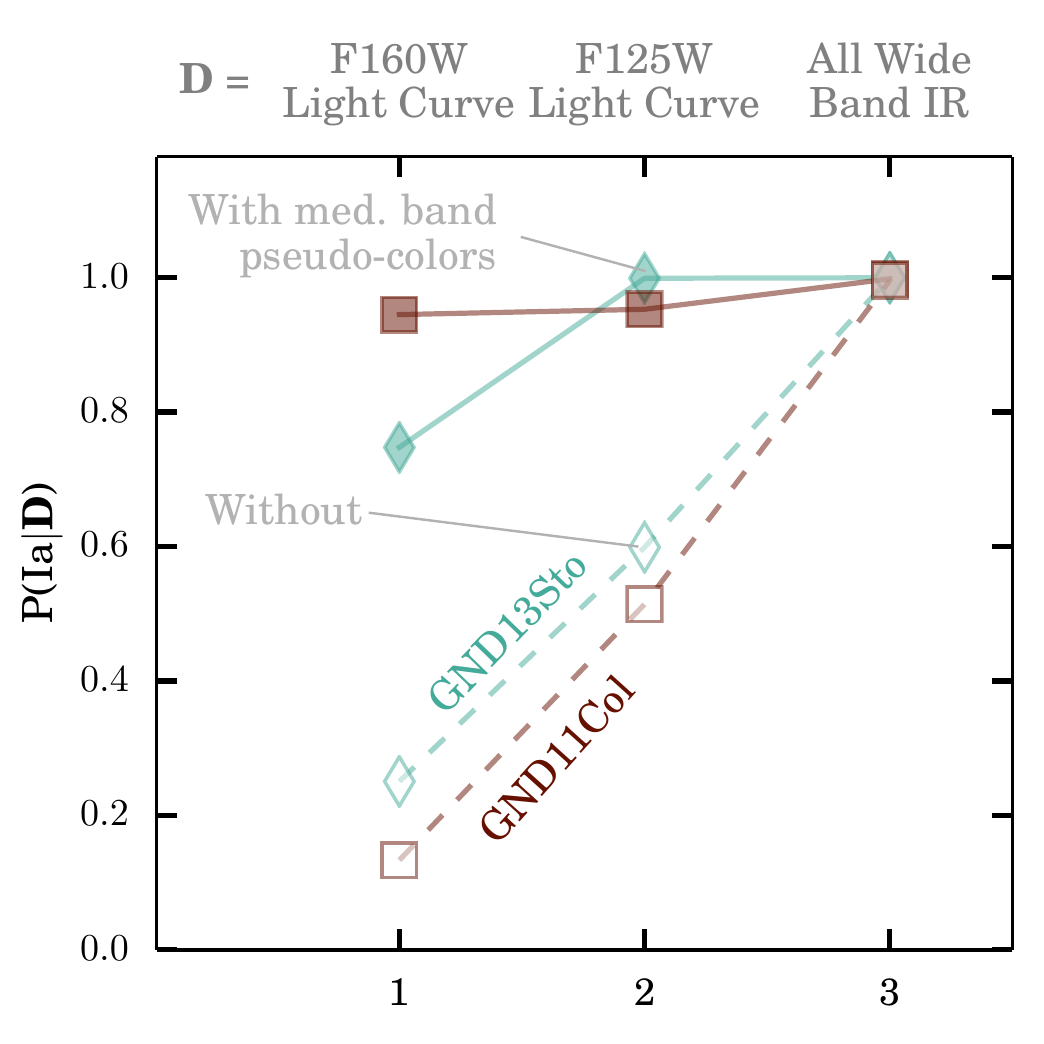}
\caption{  Examining the role of medium-band data in modifying the
Type Ia classification probability for SN \stone\ (blue diamonds)
and \colfax\ (red squares).  Along each column we plot the Type Ia
classification probability derived from a subset of the light curve
data. For column 1 we use just the F160W band light curve (with and
without a single epoch of medium-band pseudo-colors). Column 2 uses
just the F125W band light curve, and column 3 uses all available
wide-band IR data. Filled symbols with solid lines mark the results
when including medium-band imaging data, and open symbols with dashed
lines show the probabilities inferred without any medium-band data.
In all tests we include a redshift prior derived from the nearest host
galaxy candidate, but we add no prior on the time of peak brightness.
\label{fig:SubsetClasstest} }
\end{center}
\end{figure}

\changeG{
In addition to improving the SN redshifts, the medium-band data
also add value by strengthening our photometric SN
classifications.  To see how large this effect is, we have done a
piecemeal classification test: running subsets of each SN light curve
through our Bayesian photometric classification algorithm both with
and without the medium-band imaging data.
Figure~\ref{fig:SubsetClasstest} shows the results.  The value of the
medium-band data in strengthening the classification is apparent in
the first two columns of this figure.  For these two tests we used
just a single wide-band light curve (F125W or F160W) plus a single
epoch of medium-band imaging (including the necessary observations in
each complementary wide band). The addition of medium-band imaging
alters the probabilities from a misleading or ambiguous P(Ia$|${\bf
D})$\lesssim$0.5 into a much more robust P(Ia$|${\bf D})$>$0.75.  In
these cases the single-band light curve primarily constrains the
time of maximum light, while the single-epoch medium-band
pseudo-colors provides the majority of the classification
information.  In the third column of Figure~\ref{fig:SubsetClasstest},
where the full multi-band IR light curve is available, the medium-band
pseudo-colors do not alter the formal classification probability.
\change{This shows that the Type Ia classification can be securely
established without any medium-band data, but to do so requires shape
and color information afforded by multiple wide-band light curves
extending over several months.}
}

\change{
\subsection{Redshift Determination}\label{sec:RedshiftDiscussion}
}

\changeG{
In Sections~\ref{sec:StoneRedshift}
and \ref{sec:ColfaxClassificationRedshift} we saw that the medium-band
imaging \change{improves the {\it accuracy} of the final redshift
determination, by providing redshift constraints that are independent
of the presumed host galaxy. } In the case of SN \stone, where the SN
is widely separated from all nearby galaxies, the additional redshift
constraint from medium-band pseudo-colors is especially valuable as a
check against mis-identification of the host galaxy.  For SN \colfax\
-- where we have no spectroscopic data on the host galaxy -- the
redshift information derived from medium-band imaging is very valuable
to guard against a catastrophic error in the host galaxy
photo-$z$, \change{or misassociation with a galaxy that is actually in
the foreground or background.}  
}

\changeG{
\change{
Figures~\ref{fig:stonepostz} and \ref{fig:colfaxpostz} also demonstrate the
value of medium band imaging to improve the {\it precision} of
redshift constraints.  For both \stone\ and \colfax\ the medium-band
pseudo-colors deliver posterior redshift distributions with multiple
peaks.  As the broad single peak from the wide-band light curve breaks
the degeneracy by picking out a single peak, the combination of
wide- and medium-band imaging can reduce the photometric redshift
uncertainty by almost a factor of 2.  Redshifts measured from
the host galaxy can improve the precision still further, but if either of these SNe had
no apparent host galaxy, then the medium band imaging 
would have been critical to make the SN useful for cosmology.}
}

\begin{figure*}
\begin{center}
\includegraphics[width=\textwidth]{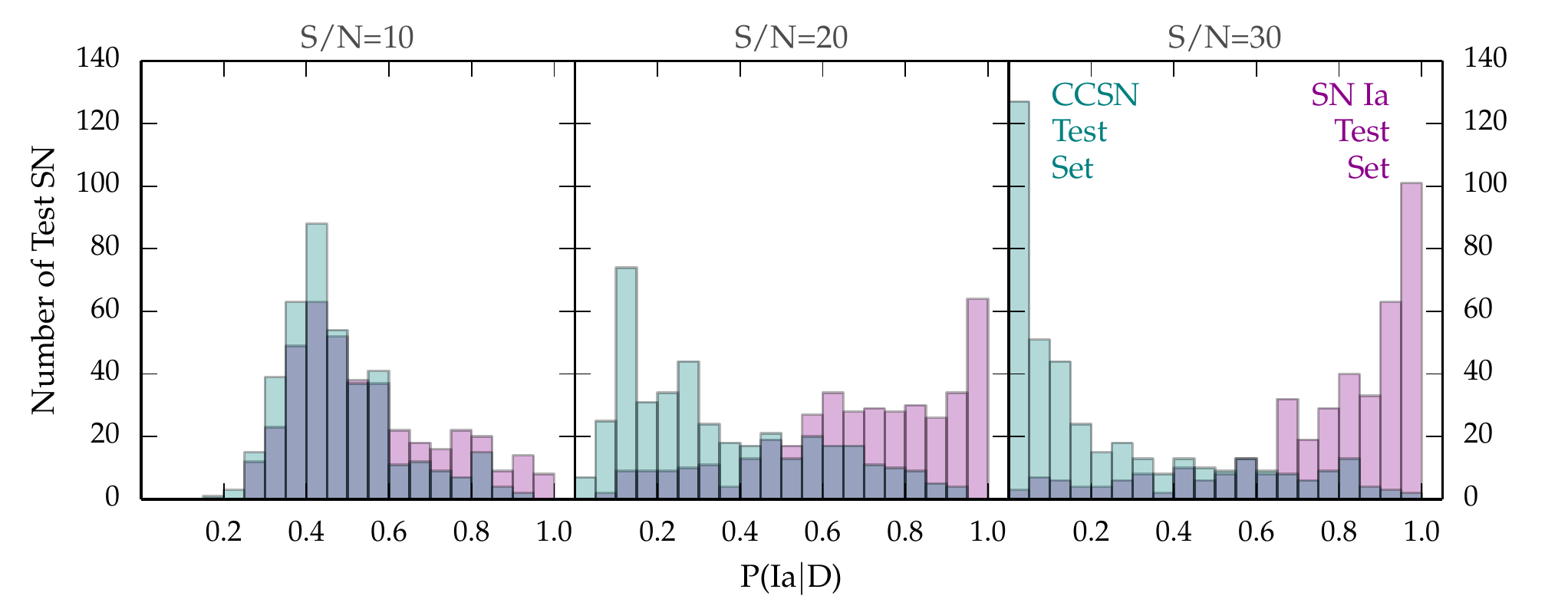}
\caption{  \label{fig:ClassTestHist}
 Results of a Monte Carlo simulation test examining the use of
 medium-band pseudo-colors alone for SN classification.  Histograms
 show the distribution of Type Ia classification probabilities derived
 by classifying 2500 synthetic SN "test particles" using {\it only} a
 medium-band pseudo-colors.  All SNe are simulated at redshifts in the
 range [1.6,2.4].  A single epoch of simulated observations are made
 in three medium band filters (F127M, F139M, F153M) and their
 corresponding wide-band counterparts (F125W, F140W, F160W), taken at
 $\pm$15 observer-frame days from peak brightness.  The blue
 histograms show classification results for simulated CC SNe and the
 red histograms count probabilities for synthetic Type Ia SNe.  The
 test particles were simulated in three rounds, with the flux
 uncertainty in every band fixed to a S/N ratio of 10, 20, or 30 --
 shown in three panels from left to right.  }
\end{center}
\end{figure*}

\change{
\subsection{Future Surveys}\label{sec:FutureSurveys}
}

\changeG{
For SN \stone\ we collected exposures totaling $\sim$5 ksec (2 HST
orbits) for each of 2 medium band filters (F139M and F153M).
For \colfax\ we used roughly 8 ksec (3 HST orbits) on each of 3
medium-band filters.  In both cases we achieved a final S/N ratio
between 5 and 10 in each individual filter, and we have seen that this
is insufficient to provide a definitive classification by itself
(e.g. Figures~\ref{fig:StoneCircle} and \ref{fig:ColfaxCircle}).  To
evaluate medium-band imaging for future surveys, it is instructive to
consider how much exposure time would have been required to make the
medium-band imaging pseudo-colors stand alone and deliver a robust
classification without reference to the wide-band light curve at all.  
}

\changeG{
To that end, we performed a Monte Carlo classification test in which
we simulated medium-band observations for 2500 synthetic SNe spread
evenly across the primary SN sub-classes, with redshifts between 1.6
and 2.4.  Each simulated SN was
``observed'' through 3 medium bands and their complementary wide-band
filters, with all observations taken concurrently in a single epoch
within $\pm$15 observer-frame days of peak brightness.  We then derive
a Bayesian classification probability P(Ia$|${\bf D}), taking care to
avoid self-identification, where a synthetic SN yields an artificially
good match when compared with the template that was used to create
it. For \CCSN\ test particles, this means that we remove from the
template library the single template used to generate the \CCSN\ test
particle. For \SNIa\ test particles, we exclude from our comparison
the range of SALT2 light curve shape and color parameters that are
similar to the values generated for the test object.  Here,
``similar'' is arbitrarily defined as a circle in ($x1$, $c$) phase
space centered on the test particle's ($x1$, $c$) values, with radius
0.05.
}

\changeG{
Figure~\ref{fig:ClassTestHist} shows histograms of the inferred type
Ia classification probabilities, derived from a series of three Monte Carlo simulations.  In
each simulation we fixed the S/N to be 10, 20 or 30 for every
individual filter.   The progression of three panels in
Figure~\ref{fig:ClassTestHist} demonstrates the result that is to be
naturally expected: as S/N is increased, the prospect of
classification from medium-band data alone becomes
more tenable.   With S/N=10 in each band, the two simulated
populations are indistinguishable.  At S/N=20 the majority of
synthetic Type Ia SNe are recovered, with P(Ia$|${\bf D})$>$0.75 -- but
there is still substantial contamination from CC SNe being incorrectly
classified.  A large Type Ia sample defined using medium-band observations
at S/N=20 would have a purity of $\sim$80\%.    When all observations
attain S/N=30, the populations are much more distinct, and it would be
possible to define a \SNIa\ sample with $>$95\%\ purity using such
high-quality medium-band data.    
}

\changeG{
Realistically, these S/N targets are not achievable with HST for
targets like SN \stone\ and \colfax.  To reach S/N=20 for a Type Ia SN
at z=2.0 with an F160W AB mag of 26 would require a total of roughly
100 ksec (40 HST orbits) spread across three medium bands and their
wide-band counterparts.  For S/N=30, the exposure time requirement
rises to $\sim$215 ksec (nearly 90 HST orbits).  At those costs, grism
spectroscopy may become an appealing alternative, especially since it
delivers more ancillary science benefits in any given field.  Given
the limitations of finite observing time, the best strategy for HST
medium-band imaging may be the one used in this work: relatively short
observations that are not definitive in themselves, but provide
support for a holistic approach to photometric classification and
redshift determination.
}

\changeG{
The small set of medium-band IR filters on HST currently limits the
application of the medium-band pseudo-spectroscopy method to SNe at
$z\sim2$. The JWST Near-infrared Camera (NIRCAM) will have 12
medium-band filters from 1.5 to 5 $\mu$m, substantially expanding the
reach of this technique and the potential for testing models of time
variable dark energy or \SNIa\ evolution.  In principle, JWST medium
band filters could be used to characterize \SNIa\ candidates as far as
$z\sim6$.  Unfortunately, the diminishing cosmic star formation rate
density at very high redshifts will make these objects few and far
between, so the initial discovery will be more efficiently done with a
wide-field observatory, such as Euclid or WFIRST-AFTA.  With this
combination of Euclid/WFIRST and JWST, it will be possible to discover
and characterize a sample of {\it primordial} \SNeIa -- explosions of
the very first generation of binary white dwarfs -- which in turn
would enable the most stringent possible test for the secular
evolution of \SNIa\ properties.
}

\change{
\section{The High-$z$ Frontier \\for Type Ia SN Cosmology}\label{sec:Cosmology}
}

\change{
The two objects presented here are among the most distant Type Ia SNe
known, with \colfax\ now setting the bar with the highest redshift.
As such, these SNe are of great interest for testing cosmological
models and evaluating possible systematic biases that may be
suppressed or obscured at lower redshifts. In this section we begin
with an extension of the Hubble diagram in
Section~\ref{sec:ExtendingTheHubbleDiagram}, and then examine some of
the particular considerations that must be addressed when
incorporating such high-$z$ SNe into a cosmological sample.}

\changeG{
\subsection{Extending the Hubble Diagram}\label{sec:ExtendingTheHubbleDiagram}
}

\begin{figure*}
\begin{center}
\includegraphics[width=\textwidth]{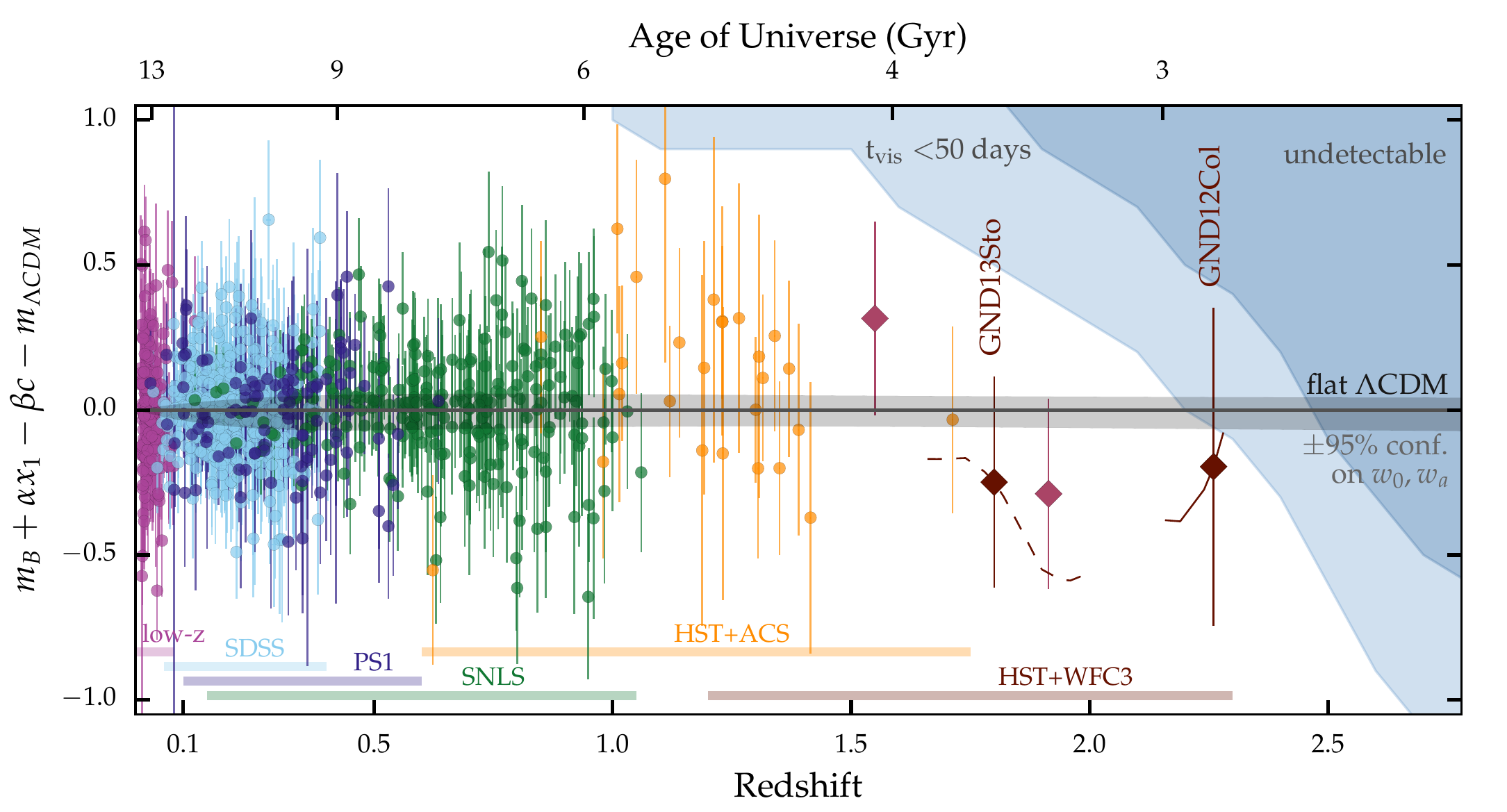}
\caption{  
Hubble residuals diagram, plotting as a function of redshift the
apparent magnitude (corrected for light curve shape and color)
relative to the predicted magnitude from a flat \LCDM\ cosmology.
Colored circles show a compilation of $\sim$900 SNe from the
low-redshift CfA and CSP
surveys \citep{Hicken:2009a,Stritzinger:2011},
SDSS \citep{Kessler:2009b}, PS1 \citep{Rest:2014},
SNLS \citep{Conley:2011}, and at $z>1$ the orange circles are from the
GOODS and SCP surveys using the ACS camera on
HST \citep{Riess:2007,Suzuki:2012}.  Red diamonds at z=1.55 and 1.91
plot two previously published \SNeIa\ from the CANDELS
survey \citep{Rodney:2012,Jones:2013}, and dark red diamonds at
$z=1.8$ and 2.26 show the two new \SNeIa\ from this work. All vertical
error bars include a uniform intrinsic dispersion of $\sigma_{\rm
int}=0.12$ mag and a redshift-dependent gravitational lensing
dispersion of $\sigma_{\rm GL}=0.093z$.  For SN \change{\stone\
and} \colfax\ the x-axis error bar is shown as a curved line to
indicate the covariance of distance with redshift for this
object \change{(dashed for \stone\ to indicate that this redshift
range only applies if the association with host galaxy A is presumed
to be spurious).} The shaded region in the upper right demarcates the space
where the CANDELS survey starts to lose sensitivity for discovery
of \SNeIa\ (see text for details).
\label{fig:hubblefig} }
\end{center}
\end{figure*}

\changeG{
In Figure~\ref{fig:hubblefig} we plot a Hubble residuals diagram,
showing the luminosity distance relative to \LCDM\ versus redshift.
The baseline cosmology is defined by fitting to a collection of
943 \SNeIa\ from the
literature \citep{Riess:1999,Jha:2006,Hicken:2009a,Kessler:2009b,
Contreras:2010,Folatelli:2010,Conley:2011,Stritzinger:2011,
Hicken:2012,Ganeshalingam:2013,Rest:2014}.
We fit these light curves using the SALT2 model, and define a
corrected apparent B-band peak magnitude following \citet{Tripp:1998}:
$m_{\rm B}^\prime = m_{\rm B} + \alpha x_1 - \beta c$.  Here the
(uncorrected) peak apparent magnitude $m_{\rm B}$, the light curve
width $x_1$, and the color $c$ are derived from the light curve fit.
For this exercise we fixed the SALT2 model parameters to the values
derived in the \citet{Rest:2014} analysis of the PS1 SN sample:
$\alpha=0.147$ and $\beta=3.13$.
We then subtract $m_{\Lambda CDM}$, the apparent magnitude predicted
by a flat \LCDM\ cosmology at each redshift, with the \SNIa\ peak
absolute magnitude at a fiducial value of $M_B=-19.34$, and using
best-fit values of \Om=0.284 and \Ho=70.88.  The SN magnitude errors
shown in Figure~\ref{fig:hubblefig} also include an intrinsic scatter
of $\sigma_{\rm int}=0.12$
magnitudes \citep{Scolnic:2014b,Betoule:2014}, and for simplicity this
term is not separately defined for each survey. These simplifications
allow for an expedient comparison of the new high-$z$ \SNeIa\ against
existing SN and a baseline cosmological model.  A more complete
examination of cosmological models will be presented in a subsequent
paper (Riess et al., in preparation).
}

\subsection{Gravitational Lensing}\label{sec:GravitationalLensing}

As \SNIa\ observations are pushed to higher redshifts, observed SN
fluxes are increasingly affected by gravitational lensing.  Photons in
the Universe are conserved, so when sampling a sufficiently large
number of random sight lines, the mean SN flux is unperturbed. Thus,
lensing adds scatter to \SNIa\ distances, but if treated appropriately
it does not bias cosmological inferences \citep{Holz:1998b}.

Lensing can be handled in one of two ways: either (a) rely on a large
sample to average out the bias, while accounting for the increased
scatter, or (b) estimate the (de)magnification along every line of
sight and include appropriate corrections.  \citet{Holz:2005} examined
the former option and found that lensing introduces an additional
redshift-dependent dispersion in the observed \SNIa\ fluxes of
$\sigma_{\rm lensing}=0.088z$ (or 0.093$z$ if treating
intrinsic \SNIa\ dispersion as Gaussian in magnitude).

A formalism for individual lensing corrections (option b) was
presented in \citet{Gunnarsson:2006}, in which galaxy detections from
deep imaging surveys are used to model dark matter halos that intersect
each SN sight line.  \citet{Jonsson:2006} applied this method to 33
SNe from the HST GOODS survey \citep{Strolger:2004,Riess:2004b} and
demonstrated that such lensing corrections can indeed reduce the
lensing scatter.  Additionally, \citeauthor{Jonsson:2006}\ found that the mean
magnification factor in the HST sample was consistent with unity,
implying that there is no evidence for a bias toward including highly
magnified objects in the high-$z$ SN cosmology sample.

In this work we have found no evidence for significant lensing from
the nearest detected foreground galaxy to each SN
(Sections~\ref{sec:StoneLensing}
and \ref{sec:ColfaxLensing}).  \change{However, the nearest foreground
galaxy is not the only possible source for gravitational lensing
perturbations.  For a more precise lensing correction, one should}
include all foreground sources out to
$\sim$1\arcmin \citep{Gunnarsson:2006,Jonsson:2006}.

As we are not pursuing a measurement of cosmological parameters in
this work, we adopt option (a) from above, applying the 0.093$z$ mag
additional scatter from \citet{Holz:2005} as an additional noise term
for all SN shown in Figure~\ref{fig:hubblefig}.  For our two
high-$z$ \SNeIa\ this corresponds to 0.17 mag at the redshift
of \stone\ and 0.2\,mag for \colfax.  \change{It is noteworthy that in
this high redshift regime we have transitioned to a point where the
scatter from gravitational lensing along the line of sight is larger
than the intrinsic scatter of the Type Ia SN population
($\sim0.12$\,mag)--though for these individual objects both of those
error terms are dwarfed by the observational uncertainties.}

\subsection{Biases and Evolution}\label{sec:BiasDiscussion}

At the high-$z$ edge of any survey, a selection bias may be introduced
as intrinsically faint objects are more likely to slip below the
detection threshold and be missed \citep{Malmquist:1920}.  The blue
shaded regions in the upper right corner of Figure~\ref{fig:hubblefig}
show the redshift and luminosity regime where such a bias could begin
to remove \SNeIa\ from our sample.  These regions are defined by
simulating a fiducial \SNIa\ ($x_1=0$, $c=0$) at each redshift and
measuring the {\it visibility time}, $t_{\rm vis}$: the number of
observer-frame days where the SN is brighter than the 50\% detection
efficiency threshold of the CANDELS SN search \citepalias[26.5 AB mag
in F125W and F160w, see][]{Rodney:2014}.  In the light blue region,
our fiducial \SNIa\ model has a visibility time shorter than the
typical cadence of the CANDELS survey (50 days).  In the dark blue
region, the visibility time drops below zero, meaning that
normal \SNeIa\ are effectively undetectable.  This does not paint a
complete picture of the Malmquist bias, as these regions depend on the
light curve shape and color, but this illustration shows that the
absence of positive Hubble residuals at $z>1.5$ should be
unsurprising, given the measured detection limits of this survey.

In a subsequent paper, we will present distances for the complete
sample of high-$z$ \SNeIa\ from CANDELS and CLASH, comprising
$\sim$15 \SNeIa\ at $1<z<1.7$ (Riess et al. in preparation).  Further
high-$z$ \SNeIa\ will be delivered by ongoing HST surveys such as the
{\it FrontierSN} program (PI:Rodney) and the {\it See Change} survey
(PI:Perlmutter).  These extensions of the \SNIa\ sample at $z\sim1.5$
will improve the sensitivity of \SNIa\ cosmology, especially in
testing for a time-variable dark energy equation of state (the ratio
of pressure to density, $w=P/\rho$). \citet{Salzano:2013} quantified
how much these and future surveys can be expected to refine our
constraints on $w$.  Adopting a common parameterization that is linear
with the scale factor,
$w=w_0+w_a(1-a)$ \citep{Chevallier:2001,Linder:2003}, \citeauthor{Salzano:2013}
found that a sample of $\sim30$ \SNeIa\ at $z>1$ should reduce the
uncertainty on $w_a$ by $\sim20\%$.

It is naturally expected that the {\it demographics} of the
observed \SNIa\ population should change with redshift, as
brighter \SNeIa\ with slower light curves are more prevalent in
star-forming galaxies, which dominate the universe at
$z>1$ \citet{Howell:2007}.  Such a shift in the composition of
the \SNIa\ population should not by itself introduce any cosmological
bias.  However, it is also possible that differences in the progenitor
systems for high $z$ \SNeIa\ could lead to luminosity biases --
deviations from the expected standard candle relations between shape,
color and luminosity. \change{Indeed, one of the original motivations
for this high redshift SN observing program was to detect SNe in the
early universe where such effects could be disentangled from
cosmological changes \citep{Riess:2006}.}

Using the full CANDELS and CLASH \SNIa\
sample, \citetalias{Rodney:2014} found evidence that the fraction
of \SNeIa\ exploding within 500 Myr after formation is $\lesssim 50\%$.
This means that a substantial fraction of all the \SNeIa\ that we might
observe at $z\sim2$ could have been formed when the universe was $<2$
Gyr old, perhaps in very low metallicity environments. The mean mass
of a \SNIa\ progenitor star must also increase with redshift, as
slow-evolving low-mass progenitors have less time available to leave
the main sequence and start accreting mass as a white dwarf. It is
conceivable that such changes in the progenitor mass or metallicity
could lead to a systematic shift in the peak magnitudes that is not
correlated with the light curve width or
color \citep{Dominguez:2001,Timmes:2003,Riess:2006}.  Finding
evidence for this effect is more challenging than improving
constraints on variable dark energy models.  \citet{Salzano:2013}
found that testing plausible models for non-cosmological evolution of
the \SNIa\ population requires $\sim$50 SNe at $1.5<z<3.5$, which
could first be done with the {\it James Webb Space Telescope} (JWST),
scheduled for launch in 2018.

\change{
\section{Summary} \label{sec:Summary}
}

\change{
In this work we have presented two primary results, where the
first is methodological and the second observational:
}

\change{
\begin{enumerate} 
\item a description of the use of HST medium-band IR 
      filters for analyzing high-$z$ SNe; and  
\item the discovery, classification, and redshift determination for two of the 
      most distant Type Ia SNe yet known.
\end{enumerate}
}

\change{
We have demonstrated that the medium-band technique can provide an
alternative to costly spectroscopic observations of high-$z$ SNe. This
method capitalizes on the semi-periodic nature of the broad absorption
features in Type Ia SN SEDs and provides a {\it photometric} tool for
examining {\it spectroscopic} features.  For classifying and
determining redshifts of high-$z$ SNe, this technique is more
efficient than spectroscopy but more precise than photometry.  By
relying on difference imaging, we avoid contamination from host galaxy
light that can plague SN spectroscopy. By using medium-minus-wide
pseudo-colors, this approach does not sample the broad-band SED colors
and is therefore insensitive to dust extinction.}

\change{For the second component of this work, we have marshaled all available
wide- and medium-band photometry and host galaxy information to
determine that SN \stone\ and \colfax\ are both Type Ia SNe at
$z\sim2$.  These objects push the frontier for Type Ia SN cosmology
into an era when the Universe was only 3 Gyr old.  Although the
uncertainties are large, we find that both are fully consistent with a
standard flat \LCDM\ cosmology, showing no obvious deviations that
might arise from extreme evolution of the Type Ia SN progenitor
population.  We leave a more robust examination of cosmological models
to future work, which will incorporate the full CANDELS and CLASH SN
samples (Riess et al. in preparation).  }

\change{
Whereas the SN samples at $z<1$ are moving quickly toward the ``big
data'' era, the high-$z$ regime is still very much limited by the
sample size.  With these high-z objects at the redshift frontier, each
requires careful attention to accommodate the unique nuances of
classification and redshift estimation.  Infrared imaging surveys from
HST, JWST and successor observatories should continue to expand the
sample of such high-$z$ SNe in coming years. A decade from now the
high-$z$ regime may well look like the $z<1$ samples of today, with
hundreds or thousands of well-sampled Type Ia SNe from observatories
such as NASA's planned WFIRST-AFTA.  That will open up the exciting
possibility of exploring a wider range of cosmological models and
evolution scenarios with these early-universe SNe.  }

\bigskip

{\bf Acknowledgments:}

We thank the anonymous referee for a patient, careful evaluation and
for helpful comments on the manuscript.  It is our pleasure to thank
program coordinators Patricia Royle and Beth Perriello, as well as the
entire STScI scheduling team for their tireless efforts that made the
CANDELS survey and the SN follow-up program possible.  We also thank
the ``CANDELS builders'': Norman Grogin, Dale Kocevski, Anton
Koekemoer, Sandy Faber and Harry Ferguson, for their substantial
investments in the design and implementation of the CANDELS
program. We further acknowledge the work of the CANDELS/CLASH SN
search teams, notably Azalee Bostroem, Tomas Dahlen, and Anton
Koekemoer, who discovered the SNe presented here in visual searches of
the HST data.  We also thank Mark Dickinson for assistance in
evaluating spectroscopic constraints on host galaxy redshifts, and
Alexandra Pope for sharing information on the Spitzer IRS spectrum of
the \stone\ host galaxy.  We thank Chuck Keeton for assistance with
GRAVLENS.

This work was principally based on observations made with the NASA/ESA
Hubble Space Telescope, which is operated by the Association of
Universities for Research in Astronomy, Inc., under NASA contract NAS
5-26555. These observations are associated with program IDs 12060,
12061, 12062, 12442, 12443, 12444, 12445, 12099, 12461, and 13063.
The analysis presented here made extensive use of the Mikulski Archive
for Space Telescopes (MAST). STScI is operated by the Association of
Universities for Research in Astronomy, Inc., under NASA contract
NAS5-26555. Support for MAST for non-HST data is provided by the NASA
Office of Space Science via grant NNX13AC07G and by other grants and
contracts.  Financial support was provided by NASA to SAR through
grant HST-HF-51312, and to the CANDELS-SN team through grants
HST-GO-12060 and HST-GO-12099 from the Space Telescope Science
Institute.

{\it Facilities:} \facility{HST (WFC3)}
\smallskip

\appendix
\section{\bush}  \label{sec:Bush}

\begin{figure*}
\begin{center}
\includegraphics[width=0.37\textwidth]{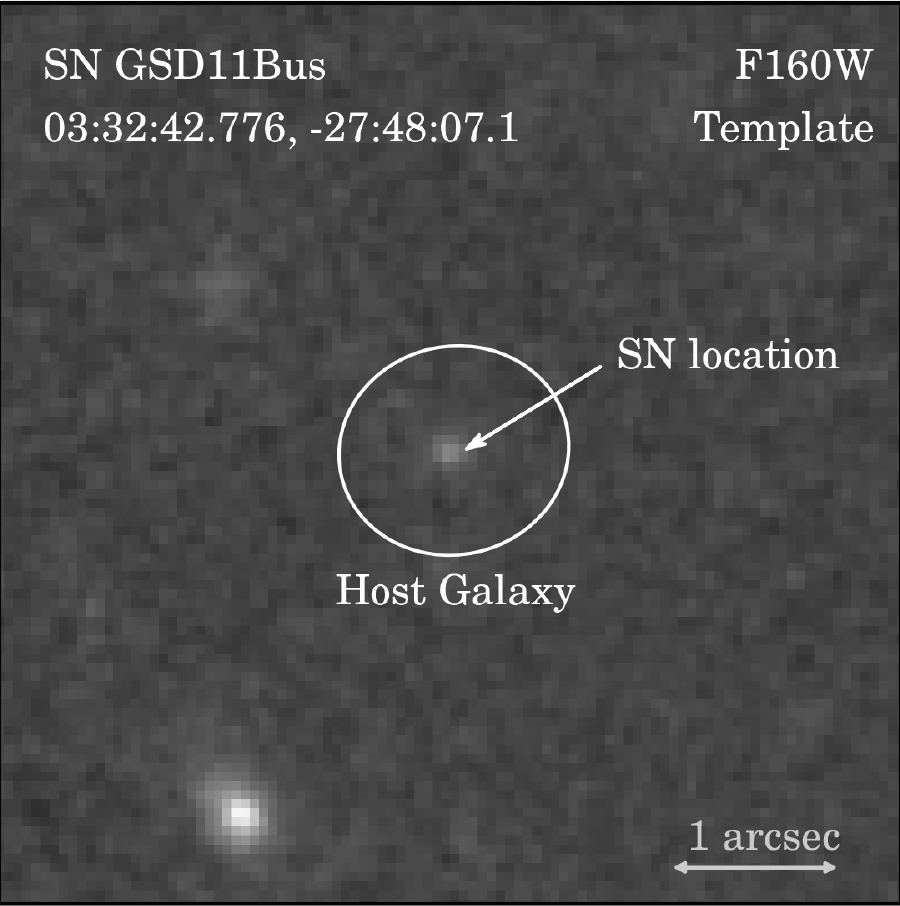}
\includegraphics[width=0.62\textwidth]{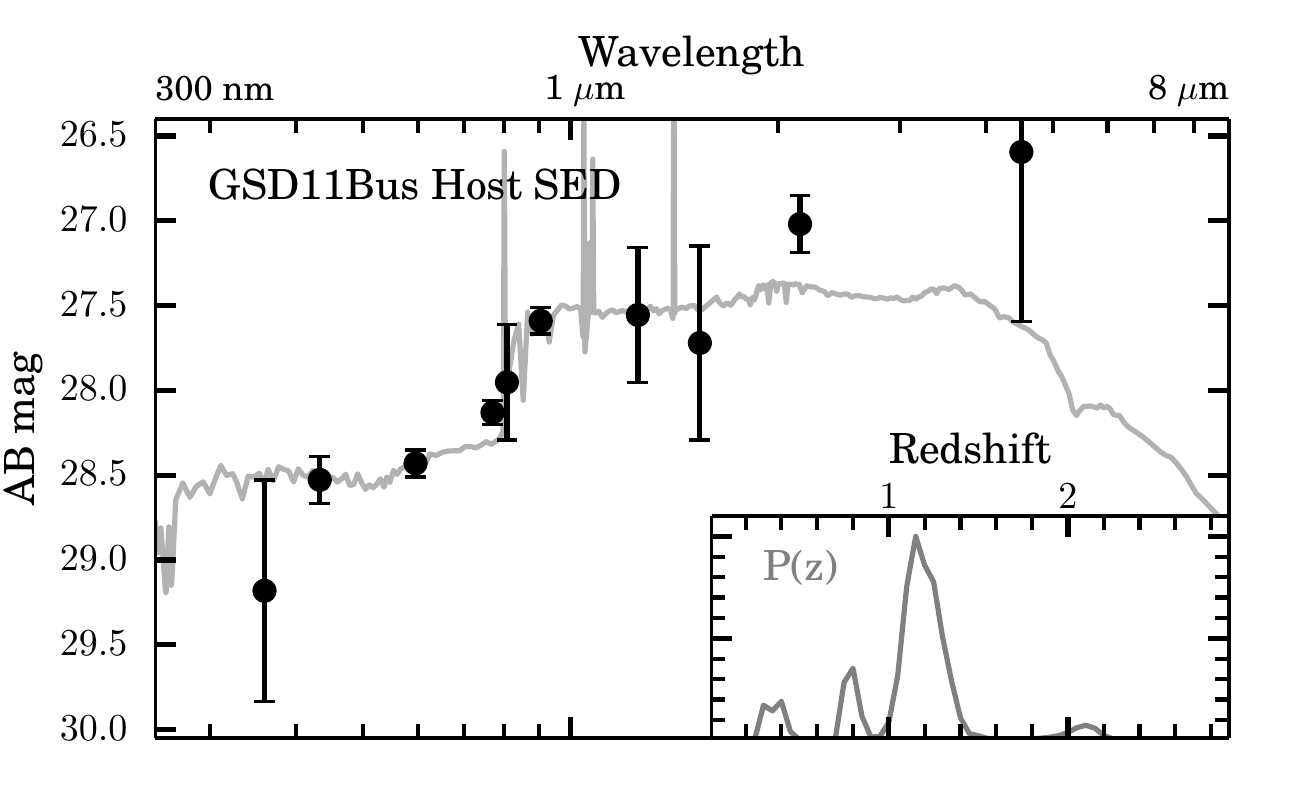}
\caption{  
The host galaxy of \bush.  {\it Left:} The F160W template image,
constructed from all exposures before MJD=55720 and after 55600. The
ellipse encloses $\sim$90\% of the host galaxy flux in the F160W band.  {\it
Right:} Filled points show the observed SED of the \bush\ host galaxy
and the gray curve is the best-fitting SED model. The inset shows the
photometric redshift probability distribution, peaked at z=1.15.
\label{fig:bushhost} }
\end{center}
\end{figure*}

\begin{deluxetable*}{rrrrrrrrr} 
\tablecolumns{9}
\tablecaption{\bush\ Photometry\label{tab:bushphot}}
\tablehead{ 
    \colhead{Obs. Date}
  & \colhead{Filter}
  & \colhead{Exp. Time}
  & \colhead{Flux}
  & \colhead{Flux Err}
  & \colhead{AB Mag\tablenotemark{a}}
  & \colhead{Mag Err}
  & \colhead{AB Zero Point} 
  & \colhead{$\Delta$ZP\tablenotemark{b}} \\
    \colhead{(MJD)}
  & \colhead{}
  & \colhead{(sec)}
  & \colhead{(counts/sec)}
  & \colhead{(counts/sec)}
  & \colhead{}
  & \colhead{}
  & \colhead{} 
  & \colhead{(Vega-AB)} 
}
\startdata
55720.1  &     F606W &  5211 &     0.0623 &  0.0497  &  29.506 &     0.865  &  26.493 & -0.086\\
55723.0  &     F606W &  5211 &    -0.2223 &  0.0546  & $>$28.5 &   \nodata  &  26.493 & -0.086\\
55795.2  &     F606W &  1007 &     0.0968 &  0.1053  & $>$27.7 &   \nodata  &  26.493 & -0.086\\[1mm]
55720.2  &     F814W &  5808 &    -0.0921 &  0.0633  & $>$27.8 &   \nodata  &  25.947 & -0.424\\
55770.9  &     F814W &  1886 &     0.3049 &  0.0864  &  27.237 &     0.308  &  25.947 & -0.424\\
55795.1  &     F814W &   932 &     0.4176 &  0.1057  &  26.895 &     0.275  &  25.947 & -0.424\\
55847.7  &     F814W &   480 &    -0.4589 &  0.1879  & $>$26.6 &   \nodata  &  25.947 & -0.424\\[1mm]
55720.0  &    F850LP &  2046 &    -0.0480 &  0.0566  & $>$26.8 &   \nodata  &  24.857 & -0.519\\
55722.9  &    F850LP &  2046 &    -0.2579 &  0.0694  & $>$26.6 &   \nodata  &  24.857 & -0.519\\[1mm]
55803.2  &     F098M &  2512 &     0.4756 &  0.0766  &  26.474 &     0.175  &  25.667 & -0.562\\[1mm]
55803.9  &     F105W &  1256 &     1.2029 &  0.1465  &  26.068 &     0.132  &  26.269 & -0.645\\
55892.5  &     F105W &  8395 &     0.2638 &  0.0774  &  27.716 &     0.319  &  26.269 & -0.645\\[1mm]
55718.1  &     F125W &  1006 &     0.1360 &  0.1691  & $>$27.0 &   \nodata  &  26.230 & -0.901\\
55778.8  &     F125W &  1006 &     0.9312 &  0.1593  &  26.308 &     0.186  &  26.230 & -0.901\\
55795.1  &     F125W &  1006 &     1.0172 &  0.1556  &  26.212 &     0.166  &  26.230 & -0.901\\
55803.1  &     F125W &  1256 &     1.2903 &  0.1558  &  25.953 &     0.131  &  26.230 & -0.901\\
55819.6  &     F125W &  1006 &     0.9515 &  0.1565  &  26.284 &     0.179  &  26.230 & -0.901\\
55854.2  &     F125W &  1006 &     0.5491 &  0.1548  &  26.881 &     0.306  &  26.230 & -0.901\\
55869.9  &     F125W &  1006 &     0.3475 &  0.1442  &  27.378 &     0.451  &  26.230 & -0.901\\
55923.2  &     F125W &  1006 &     0.4099 &  0.1440  &  27.198 &     0.382  &  26.230 & -0.901\\
55974.6  &     F125W &  1006 &     0.2356 &  0.1477  &  27.800 &     0.681  &  26.230 & -0.901\\[1mm]
55803.3  &     F127M &  2512 &     0.2966 &  0.0658  &  25.961 &     0.241  &  24.641 & -0.961\\[1mm]
55803.3  &     F139M &  2512 &     0.3664 &  0.0725  &  25.569 &     0.215  &  24.479 & -1.079\\[1mm]
55803.9  &     F140W &  1256 &     1.6063 &  0.2136  &  25.937 &     0.144  &  26.452 & -1.076\\
55847.8  &     F140W &   812 &     0.9162 &  0.2536  &  26.547 &     0.300  &  26.452 & -1.076\\[1mm]
55803.8  &     F153M &  2512 &     0.3983 &  0.0776  &  25.463 &     0.211  &  24.463 & -1.254\\[1mm]
55718.1  &     F160W &  1056 &     0.0097 &  0.1468  & $>$26.8 &   \nodata  &  25.946 & -1.251\\
55778.8  &     F160W &  1006 &     0.7075 &  0.1900  &  26.322 &     0.291  &  25.946 & -1.251\\
55795.1  &     F160W &  1006 &     1.0666 &  0.1808  &  25.876 &     0.184  &  25.946 & -1.251\\
55803.1  &     F160W &  1256 &     1.0519 &  0.1437  &  25.891 &     0.148  &  25.946 & -1.251\\
55819.6  &     F160W &   503 &     0.8344 &  0.2442  &  26.143 &     0.318  &  25.946 & -1.251\\
55854.2  &     F160W &  1156 &     0.9717 &  0.1809  &  25.977 &     0.202  &  25.946 & -1.251\\
55869.9  &     F160W &  1056 &     0.5799 &  0.1598  &  26.538 &     0.299  &  25.946 & -1.251\\
55923.2  &     F160W &  1006 &     0.0633 &  0.3185  & $>$26.0 &   \nodata  &  25.946 & -1.251\\
55974.6  &     F160W &  1206 &     0.1534 &  0.1777  & $>$26.6 &   \nodata  &  25.946 & -1.251\\
\enddata
\tablenotetext{a}{For \change{flux values of less than 1$\sigma$ significance} we report the magnitude as a 3$\sigma$ upper limit}
\tablenotetext{b}{Zero point difference: the magnitude shift for conversion from AB to Vega magnitude units.}
\end{deluxetable*}

\begin{figure*}
\begin{center}
\includegraphics[width=\textwidth]{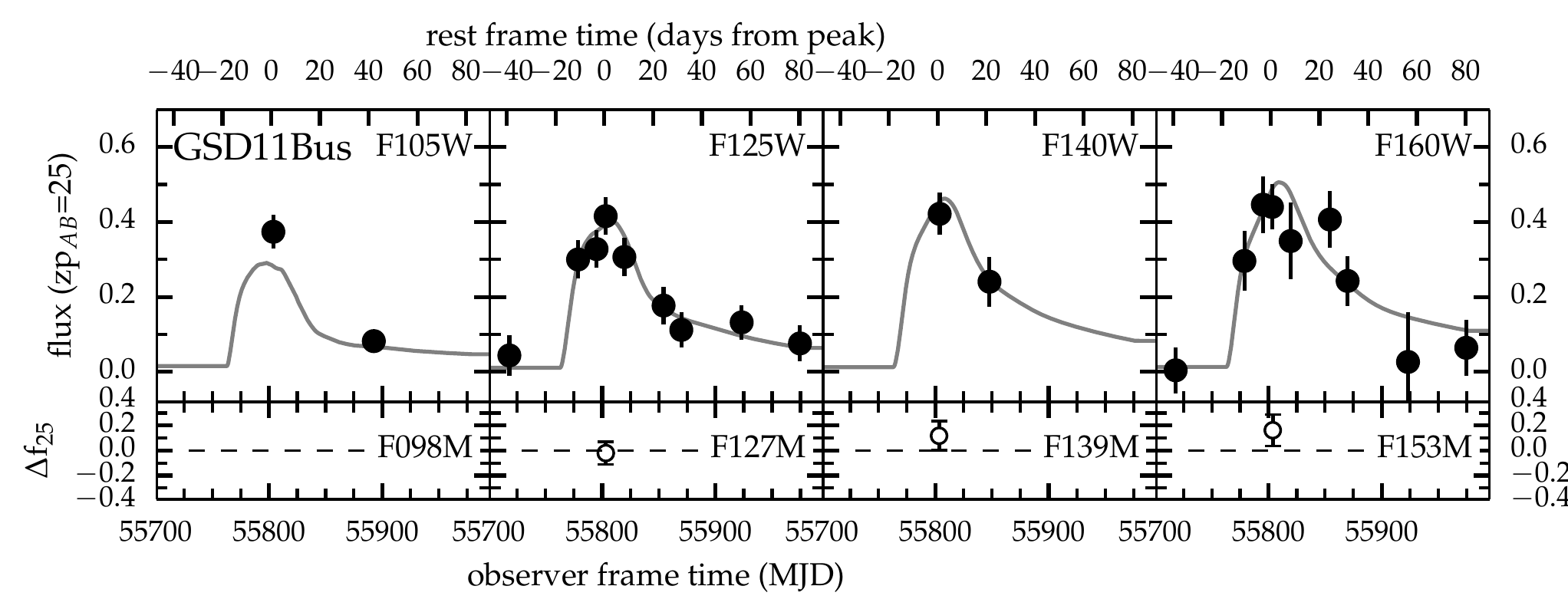}
\caption{  
Infrared light curve of \bush, showing flux as a function of time. As
in Figures~\ref{fig:stonelc} and \ref{fig:colfaxlc} fluxes are
normalized to zero point 25 and lower panels show residual fluxes for
medium-band observations. The top axis marks rest-frame time at
redshift $z=1.15$.  Grey curves show the best-fit model, based on the
Type Ic SN 2006fo.
\label{fig:bushlc} }
\end{center}
\end{figure*}

\begin{figure}
\begin{center}
\includegraphics[width=0.5\textwidth]{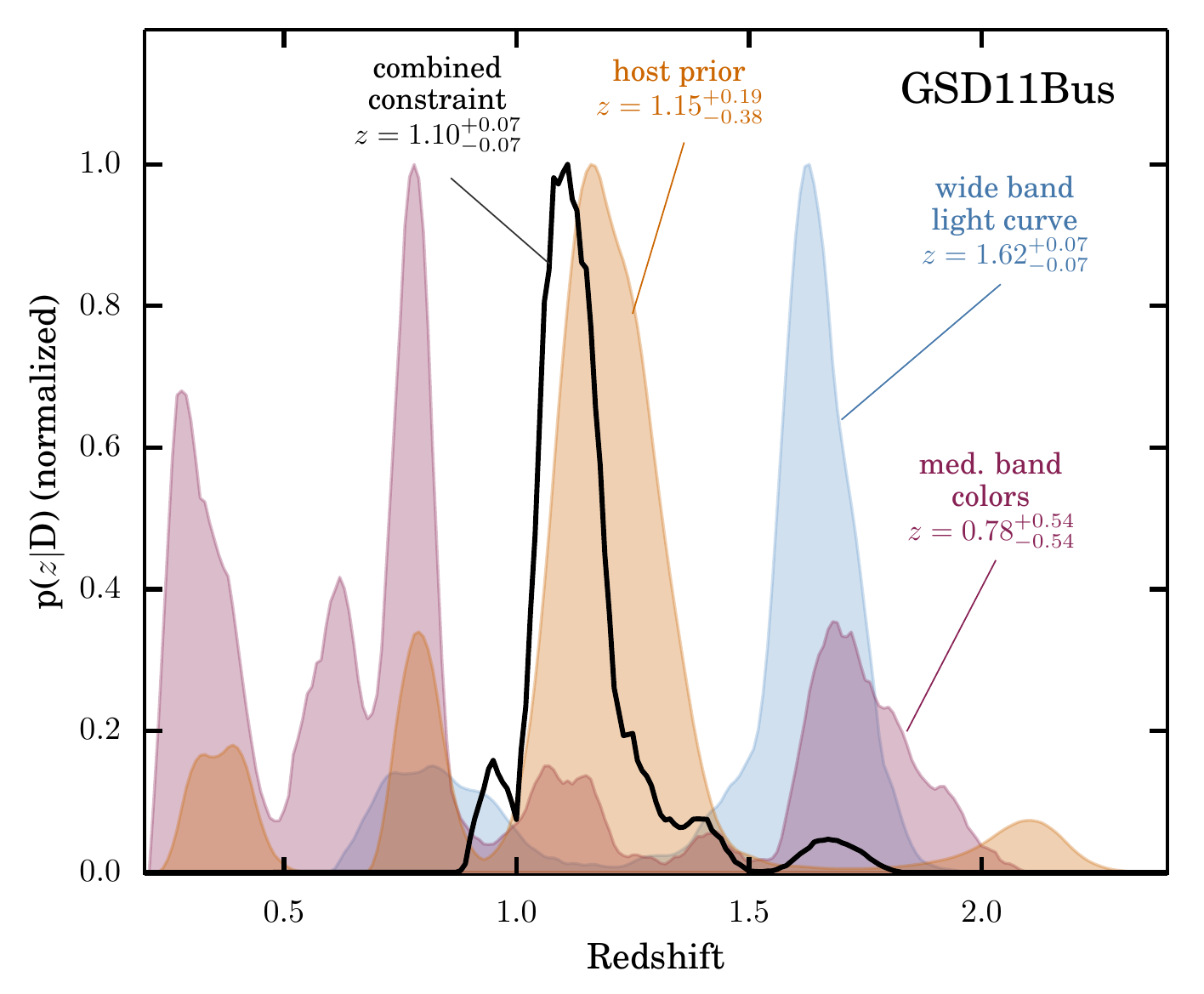}
\caption{  
Redshift constraints on \bush. As in Figures~\ref{fig:stonepostz}
and \ref{fig:colfaxpostz}, the independent constraint from each of 3
different sources and the combined constraint are shown as probability
distribution functions over redshift, all normalized to a peak value
of 1.0. }
\label{fig:bushpostz}
\end{center}
\end{figure}

\begin{figure*}
\begin{center}
\includegraphics[width=\textwidth]{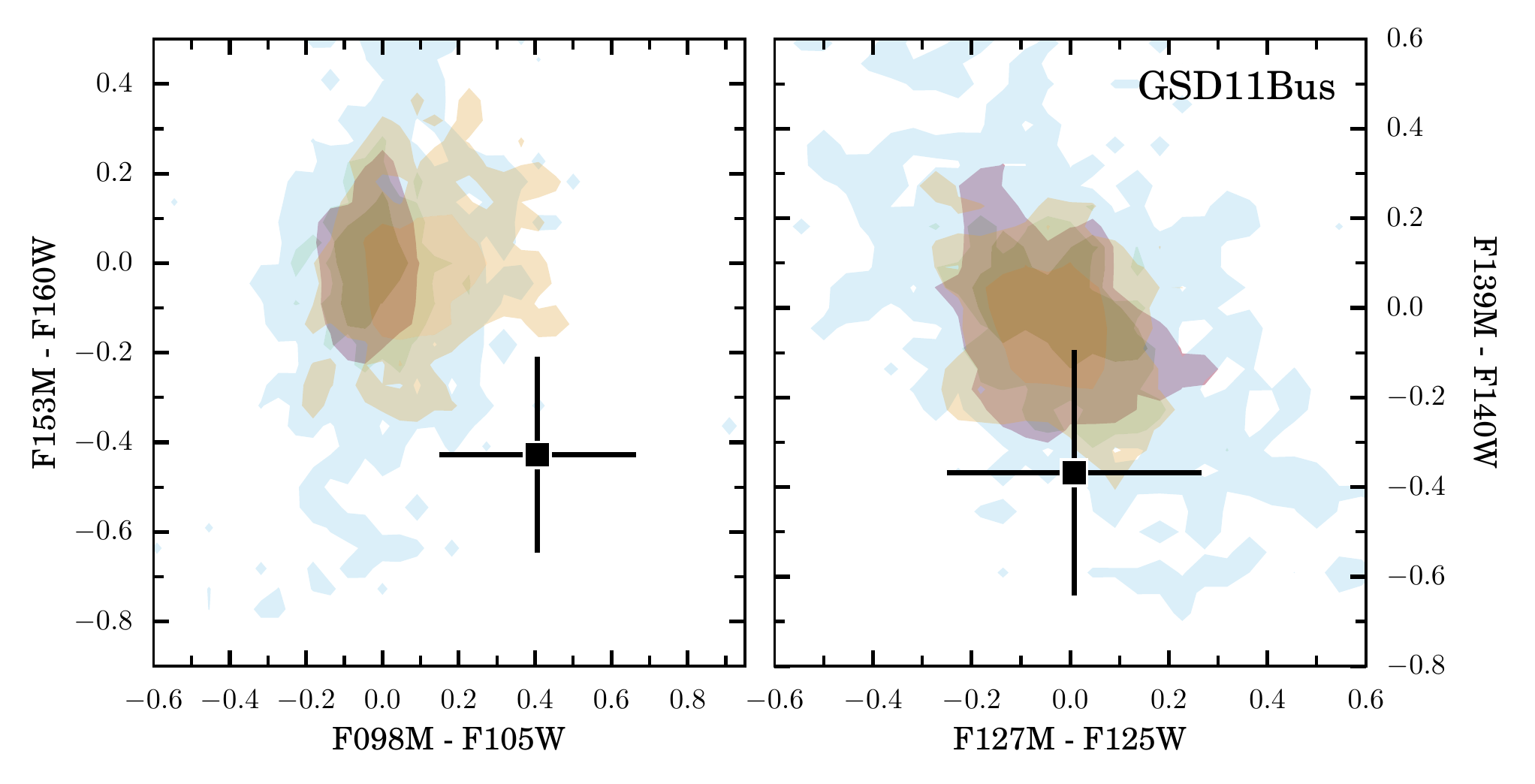}
\caption{  
Medium-band color-color diagrams for SN~\bush, as in
Figure~\ref{fig:ClassifyDemo}. Contours
enclose 68\% and 95\% of the population for each class, with red for
Type Ia, green for Type II and gold for Type Ib/c.  The simulated SN
have redshifts that evenly sample the redshift range [0.7,1.5],
encompassing the two primary peaks of the host galaxy
photo-z. Observed colors for \bush\ are indicated by the black diamond
with error bars.}
\label{fig:bushcircle}
\end{center}
\end{figure*}

\begin{figure}
\begin{center}
\includegraphics[width=0.5\textwidth]{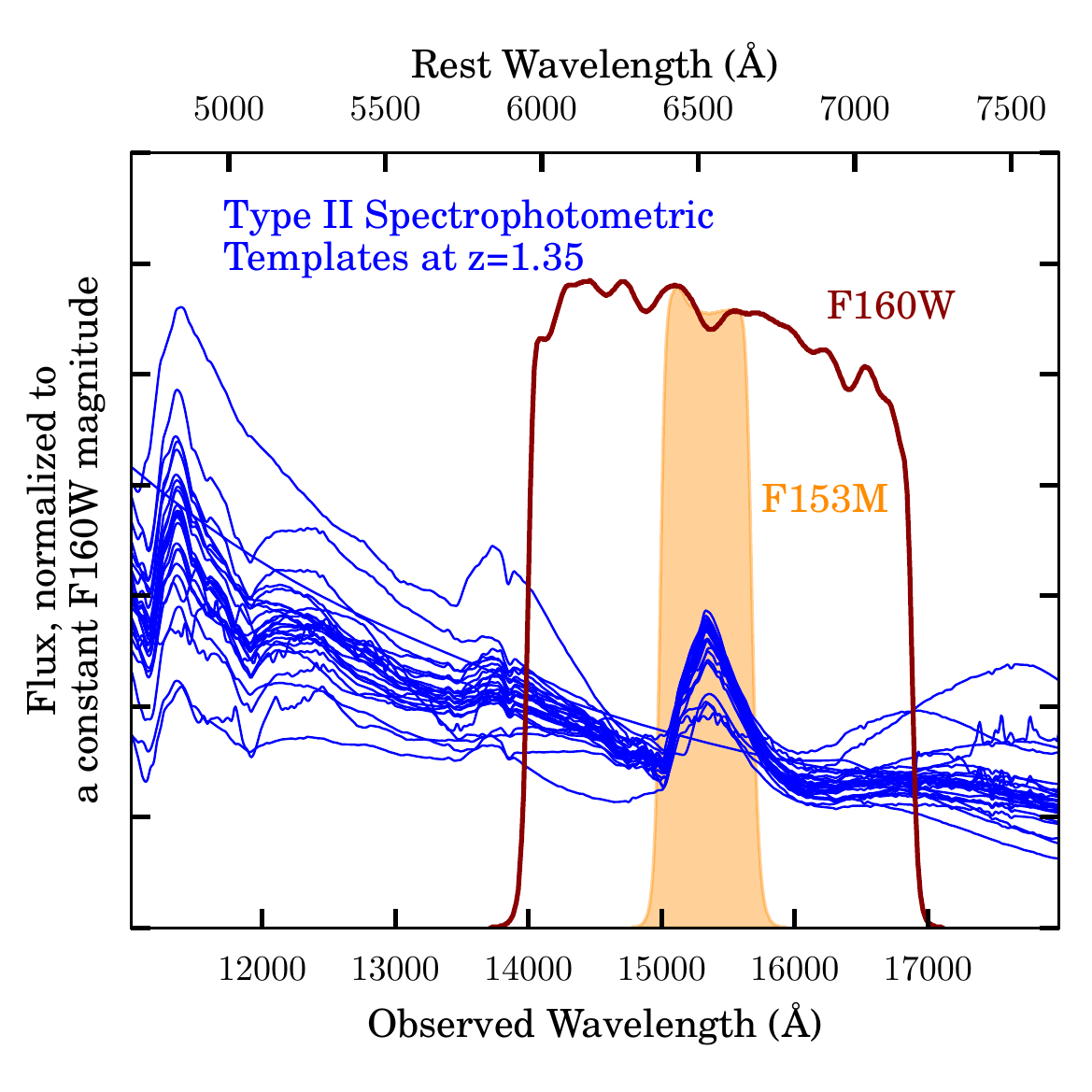}
\caption{  
Demonstrating the lack of spectral diversity in the Type II template
library employed in this work.  All Type II spectrophotometric
templates are shown as blue lines, redshifted to $z=1.35$ and observed
at +6 days after peak brightness (a redshift and age
consistent with the SN \bush\ data). Each template
spectrum is normalized to have the same integrated flux through the
F160W bandpass, which is shown as a solid red line.  At this redshift
the H$\alpha$ line occupies the entire F153M bandpass, which is shown
as the orange shaded region, but our library of models exhibits
relatively little variation of the H$\alpha$ equivalent width. 
\label{fig:TypeII_Halpha} }
\end{center}
\end{figure}

Detected in the CANDELS GOODS-S
field, \bush\footnote{\change{Nicknamed ``SN Bush'' after the 41$^{\rm st}$
and 43$^{\rm rd}$ Presidents of the United States.}} appeared in a very low
surface brightness host galaxy (26.8 AB mag arcsec$^{-2}$ in the F160W
band).  The host also has a red SED, with a V-H color of approximately
1 mag.  Identification of the 4000 \AA\ break across the {\it riz}
bands leads to a central peak in the photo-z at $z=1.15$ as shown in
Figure~\ref{fig:bushhost}.  However, the large photometric
uncertainties allow for other possible solutions at $z\sim0.3$, 0.8
and 2.1.  No spectroscopic constraint on the host redshift is
available.

The SN was initially classified as a possible SN Ia at $z>1$ based on
its very red optical-IR colors. This prompted follow-up imaging with
HST, using both medium and wide bands.  However, with the full light
curve this object was found to be inconsistent with being a normal SN
Ia anywhere in the allowed redshift range. \citetalias{Rodney:2014}
classified \bush\ as a Type Ib/c, with a redshift of $z=1.7 \pm 0.1$
based on the combination of the host photo-z and the SN light
curve.\footnote{In \citetalias{Rodney:2014} the photo-z for
the \bush\ host galaxy was reported as $z=1.76\pm0.53$, based on an
earlier version of the SED.}  Using our improved photometry
(Table~\ref{tab:bushphot}), we revisit this classification, and find
again that \bush\ is incompatible with SN Ia models. The
classification probabilities are P(Ia$|${\bf D})=0.03, P(Ib/c$|${\bf
D})=0.92, and P(II$|${\bf D})=0.05.  The best-fit light curve model
shown in Figure~\ref{fig:bushlc} is based on the Type Ic SN 2006fo
from the SDSS SN survey \citep[SDSS ID 13195,][]{Zheng:2008}.  At
$z=1.15$ this model provides a good fit to the optical+IR light curve,
with $\chi^2_{\nu}=0.85$.

Figure~\ref{fig:bushpostz} shows the redshift constraints on \bush\
from the wide-band light curve, medium-band colors, the host photo-z,
and the combined constraint from all sources.  Without the host prior,
the light curve constraints would favor a redshift at $z=1.6$ or
0.8. These redshifts also require a \CCSN\ classification, so although
the host and light curve constraints are not in agreement, there is no
plausible solution in which \bush\ can be classified as a Type Ia.
The Type Ia classification is strongly disfavored principally because
the slow decline of the light curve favors a high redshift, which is
incompatible with the observed fluxes in the ACS F814W and WFC3-IR
F105W bands.  This is reinforced by the medium-band colors, shown in
Figure~\ref{fig:bushcircle}, which locate \bush\ in a region of
pseudo-color space that is well removed from the \SNIa\ locus. 

\change{ Although this evidence is sufficient to rule out a Type Ia
classification, one may be concerned that the \bush\ medium band
pseudo-colors place it outside of the 95\% contours for {\it all} SN
classes, as shown in the left panel of Figure~\ref{fig:bushcircle}. It
is possible that \bush\ is simply a statistical outlier, but we must
caution that this discrepancy may be indicative of a fundamental
problem with the library of CC SN templates employed in this work.
Although the template library includes 26 Type II templates, these are
all based on a set of just 3 spectral time
series \citep{Gilliland:1999,DiCarlo:2002,Baron:2004}. Each template
was generated by warping those few spectral time series to match the
well-sampled broad-band photometry of 26 low-$z$ SN.  In this way, the
broad-band colors of our simulated SNe can be expected to accurately
represent the diversity of broad-band colors from this population.
However, the diversity of emission line strengths may not be well
represented in this library, as the warping procedure is not sensitive
to such features.  
}

\change{
Although we have classified \bush\ as a Type Ib/c object, the light
curve may also be consistent with the Type II-L or IIn sub-classes, which
are under-represented in our template library.  For those cases, the
strong H$\alpha$ feature seen in most Type II SNe may be particularly
relevant to SN \bush, as that feature intersects the F153M band at
redshifts near $z=1.35$, close to the peak of the \bush\ host
photo-$z$ distribution. Figure~\ref{fig:TypeII_Halpha} shows that our
template library has a marked lack of diversity in H$\alpha$ flux
relative to the integrated flux through the encompassing broad band.
Direct observations have shown a much wider range of diversity in the
strength and shape of the H$\alpha$
feature \citep{Patat:1994,Gutierrez:2014}, as would be expected for
this heterogeneous class.  This suggests that the medium-band
photometry for SN \bush\ could lead to a more definitive
classification and redshift if we had a better set of CC SN templates
that more accurately represents the true diversity of these SN
types. }

\bibliographystyle{apj}
\bibliography{bibdesk}

\end{document}